\documentclass[a4paper,fleqn,usenatbib]{mnras}

\def\aj{AJ}

\def\apj{ApJ}
\def\apjl{ApJ}

\def\aap{A\&A}

\def\mnras{MNRAS}

\usepackage{newtxtext,newtxmath}

\usepackage[T1]{fontenc}
\usepackage{ae,aecompl}
\usepackage{xfrac}
\usepackage{nicefrac}

\usepackage{graphicx}	
\usepackage{amsmath}	
\usepackage{amssymb}	
\usepackage{multirow}
\makeatletter
\newlength{\abovecaptionskip}%
\makeatother
\usepackage{threeparttable}
\usepackage{color}

\title[Revisiting the Stellar Velocity Ellipsoid -- Hubble type relation]{Revisiting the Stellar Velocity Ellipsoid -- Hubble type relation: observations versus simulations}

\author[F. Pinna et al.]{F. Pinna$^{1,2}$,\thanks{E-mail: fpinna@iac.es}
J. Falc\'on-Barroso$^{1,2}$,
M. Martig$^{3}$,
I. Mart\'inez-Valpuesta$^{1,2}$,
\newauthor
J. M\'endez-Abreu$^{1,2}$,
G. van de Ven$^{4,5}$,
R. Leaman$^{5}$,
and M. Lyubenova$^{4}$
\\\\
$^{1}$Instituto de Astrof\'isica de Canarias, C/ Via L\'actea s/n, E-38200 La Laguna, Tenerife, Spain\\
$^{2}$Depto. Astrof\'isica, Universidad de La Laguna (ULL), E-38206 La Laguna, Tenerife, Spain \\
$^{3}$Astrophysics Research Institute, Liverpool John Moores University, 146 Brownlow Hill, Liverpool L3 5RF, UK \\
$^{4}$European Southern Observatory, Karl-Schwarzschild-Str. 2, 85748 Garching b. M\"unchen, Germany \\
$^{5}$Max-Planck-Institut f\"ur Astronomie, K\"onigstuhl 17, D-69117 Heidelberg, Germany
}

\date{Accepted XXX. Received YYY; in original form ZZZ}

\pubyear{2017}

\begin{document}
\label{firstpage}
\pagerange{\pageref{firstpage}--\pageref{lastpage}}
\maketitle

\begin{abstract}
The stellar velocity ellipsoid (SVE) in galaxies can provide important 
information on the processes that participate in the dynamical heating of their 
disc components (e.g. giant molecular clouds, mergers, spiral density waves, bars). Earlier findings suggested a strong relation between the shape 
of the disc SVE and Hubble type, with later-type galaxies displaying more anisotropic 
ellipsoids and early-types being more isotropic. In this paper, we revisit the 
strength of this relation using an exhaustive compilation of observational 
results from the literature on this issue. We find no clear correlation between the 
shape of the disc SVE and morphological type, and show that galaxies with the same 
Hubble type display a wide range of vertical-to-radial velocity dispersion ratios. 
The points are distributed around a mean value and scatter of $\sigma_z/\sigma_R=0.7\pm 0.2$.
With the aid of numerical 
simulations, we argue that different mechanisms might influence the shape of the SVE 
in the same manner and that the same process (e.g. mergers) does not have the same impact in all the galaxies. The complexity of the observational picture is confirmed by 
these simulations, which suggest that the vertical-to-radial axis ratio of the
SVE is not a good indicator of the main source of disc heating. Our analysis of 
those simulations also indicates that the observed shape of the disc SVE may be 
affected by several processes simultaneously and that the signatures of 
some of them (e.g. mergers) fade over time.\looseness-2
\end{abstract}


\section{Introduction}  \label{sec1}

Ever since its discovery in the Milky Way, the relation between stellar random motions and age 
has been a matter of intense study \citep[e.g.][]{Spitzer1953, Wielen1977, Carlberg1985, Holmberg2009}. 
There are two main invoked scenarios, explaining why old stars display larger velocity dispersions than young stars. 
On the one hand, an increasing gas velocity dispersion with redshift has been observed in disc galaxies \citep{Wisnioski2015}. 
This suggests that old stars, born from this turbulent and dynamically hot gas, would have already high dispersions when they were young. 
Such cases were shown both in \citet{House2011} and \citet{Bird2013}. 
Some galaxies, in the \textit{N}-body hydrodynamical simulations sample from the former study, maintained the same dispersions until the end of the 
simulations. 
In galaxies with a highly turbulent interstellar medium (from mergers or internal processes), old stars were already dynamically hot at birth. 
The latter study analyses the Eris cosmological simulation \citep{Guedes2011} and agrees with an "upside-down" evolution for the Milky Way disc. 
The oldest stars would have been formed in an early active merger phase and quickly scattered into kinematically hot orbits by these mergers 
\citep{Bird2013}. 

On the other hand, in a disc heating scenario, stars were born in a very thin 
layer of gas with cold orbits and observed random velocities would have 
appeared more recently \citep[e.g.][]{Merrifield2001}. 
This occurs in other galaxies of the simulations sample from \citet{House2011}, where disc stars were born dynamically cold 
and acquired higher dispersions later, thanks to various disc heating mechanisms.
A number of candidate sources for dynamical heating have been 
identified. These include encounters with giant molecular clouds (GMCs), 
perturbations from irregular and transient spiral structures or from stellar 
bars, dissolution of young stellar clusters, scattering by dark halo objects or 
globular clusters, and disturbances by satellite galaxies or mergers 
\citep[e.g.][]{Gerssen2012}.\looseness-2

It is generally believed that the relative amplitudes of the random motions in different directions are a 
measure of the relative importance of the disc heating agents, and can provide a 
diagnostic for the possible sources \citep[e.g.][]{Merrifield2001}. The ellipsoidal 
three-dimensional distribution of the stellar velocity dispersions was originally 
introduced by \citet{Schwarzschild1907}. The stellar velocity 
ellipsoid (SVE), in cylindrical coordinates formed by the vertical 
($\sigma_{z}$), radial ($\sigma_{R}$) and azimuthal ($\sigma_{\phi}$) components 
of the stellar velocity dispersion, is thus the result of 
the combination of the predominant heating mechanisms. The ratios $\sigma_{z}/\sigma_{R}$, 
$\sigma_{\phi}/\sigma_{R}$ and $\sigma_{z}/\sigma_{\phi}$ define its shape. In 
an axisymmetric disc with stellar orbits not too far from circular (the 
\textit{epicycle approximation}, see \citealt{Binney1987}), 
$\sigma_{\phi}/\sigma_{R}$ depends only on the circular velocity and no 
heating mechanisms. This is the main reason why several authors have used the 
measurements of the $\sigma_{z}/\sigma_{R}$ ratio to make predictions about the 
predominant heating processes in galactic discs \citep[e.g.][]{Gerssen2012}. 
Vertical and radial dispersions similar to each other would reveal an \textit{isotropic heating}, 
whereas \textit{anisotropic heating} would be displayed by a significant difference between those velocity dispersion components.

The candidate agents for the dynamical heating of discs can be classified as:
\begin{enumerate}
\item \textbf{three-dimensional agents}, responsible for increasing both 
vertical and radial (and azimuthal) dispersions, although not necessarily at the 
same rate. The most invoked examples in the literature are encounters with 
GMCs and galaxy mergers.
\item \textbf{radial (or planar) agents}, involved only in radial (planar) 
heating, with no effect on the vertical component of the velocity dispersion. 
The most invoked candidate is the spiral structure typical of most disc 
galaxies, but bars can also have similar effects.
\end{enumerate}

\citet{Spitzer1953} showed that the star-cloud encounters, related to the 
large-scale fluctuations in the interstellar medium density, could be 
responsible for the increase of the velocity dispersion with advancing spectral 
type (and age) along the stellar main sequence. Moreover, clustered star 
formation may add kinematically hot components to pre-existing populations, 
thanks to the expulsion of the residual gas by the massive stars, thickening the 
disc \citep{Kroupa2002}. Recent \textit{N}-body simulations by \citet{Aumer2016} 
confirm that GMCs are capable of heating cold discs vertically as well as 
horizontally. If their influence was significantly higher in the past than now, 
due to higher mass and/or number densities, they could explain the observations 
in the solar neighborhood (in combination with spiral heating). Hence, their  
heating efficiency would have declined over time due to a decline in the 
star-formation rate (SFR), but also because a GMC is individually more effective 
at heating when it represents a large fraction of the not-fully-grown disc mass. 
As a consequence, we probably need to invoke other heating agents to explain the 
level of SVE isotropy observed in the later stages of the discs, when they are already massive. 

Several articles in the literature have already proposed mergers as responsible 
for enhancing $\sigma_z$ relative to $\sigma_R$. 
\citet{Toth1992} showed, with an analytical model, how the satellite infall
helps to explain the observed scale height of the Milky Way disc.
\citet{Sellwood1998}, on the 
basis of \textit{N}-body simulations, claimed that mergers with satellites 
couple closely to oscillations that can be excited in the vertical direction. 
So, direct vertical heating of disc stars by the gravitational perturbations 
would be negligible if compared to the one caused by dissipation of large-scale 
bending waves, excited by satellites' decaying orbits in resonance with stellar 
orbits. On the other hand, \citet{Benson2004} developed an analytical model, 
ignoring the effect of resonances, which predicted a median thickening of the 
discs by satellite accretions, which deposit kinetic energy into  disc stars. 
\citet{Velazquez1999}, also using \textit{N}-body simulations, studied the 
amount and distribution of the heating caused by interactions with merging 
satellites. They found that the effect depends strictly on the masses of the 
disc, the bulge and the satellites relative to each other, and on the coupling 
of their orbits with each other (prograde or retrograde).
In simulations by \citet{Kazantzidis2008,Kazantzidis2009},
vertical, radial and azimuthal velocity dispersions increased at the same time during mergers with satellites,
without affecting significantly the shape of the velocity ellipsoid.
This is in agreement with this kind of interactions being proposed as three-dimensional heating agents.

\citet{Roskar2013} claimed that the change in the radial positions of stars would have an important effect on the final disc scale height. 
On the other hand, numerous other authors showed that the contribution of this so-called radial migration might be insignificant 
\citep[e.g.][]{Minchev2012a,Minchev2012b,Minchev2012c,Minchev2013,Bird2013,Vera2014,Grand2016}.

Regarding the dynamical heating in the disc plane, numerous numerical studies 
have pointed to spiral arms as the most efficient source. Disc stars usually 
receive, from the spiral density waves, kicks at a frequency which is close to 
their natural oscillation frequency in the radial direction 
\citep[e.g.][]{Merrifield2001,Minchev2006}. Therefore, radial random motions can be increased rapidly by this near-resonant process. At the same 
time, this does not happen in the vertical direction, so that $\sigma_R$ will be 
made significantly greater than $\sigma_z$ \citep{Jenkins1990}. 
\citet{Sellwood1984} argued that the transient spiral patterns are 
self-regulating features, since they tend to stability rising secularly the 
radial velocity dispersion, until the Toomre parameter \citep{Toomre1964} has 
risen to a value around 2. This would make the spiral patterns fainter in later 
times, unless any cooling process ensures that the disc does not completely 
stabilize and exhibits a recurrent spiral structure. 

Concerning the effect of 
bars, 
\citet{Grand2016} claimed that they are important contributors to disc heating, after associating the bar buckling to a sudden increase in 
vertical kinetic energy.
\citet{Saha2010} found an explanation for the vertical heating in the inner 
region of the disc (within a scale-length) in the instabilities caused by the 
bar growth.
On the contrary,
in the outer disc, the radial heating overtook the vertical heating in presence of bars evolved to saturation 
(in the sense that its amplitude cannot grow anymore). 
\citet{Saha2013} related the buckling instability with a drop in $\sigma_z/\sigma_R$ and warned about the 
difficulty of tracking stars subject to rapid changes during the buckling phase. 

There are very limited observational works in the literature constraining the 
shape of the SVE in external galaxies since it is, still nowadays, an arduous task. One of the most 
successful studies is that of \citet{Gerssen2012}. They 
fitted the stellar velocity ellipsoids of 8 intermediately-inclined galaxies and 
studied their shapes as a function of the Hubble type. They found a strong trend 
with early-type galaxies having more isotropic SVE than later types, suggesting 
three-dimensional agents are the main contributors to the heating of their disc. 
On the contrary, in late-type discs, they proposed that  their flatter 
ellipsoids were due to the action of radial agents, most likely the spiral 
density waves. This trend was in agreement with expectations from some previous 
studies \citep[e.g.][]{Jenkins1990}, but not others 
\citep[e.g.][]{vanderKruit1999}. 
Globally, there is no consensus in the 
literature about the shape of late-type ellipsoids, in the sense that 
different results have been found for different galaxies of the same 
morphological type.
Even for our Galaxy, where the properties of individual stars can be measured directly, 
a wide range of values of the vertical-to-radial velocity dispersion ratio were observed in the solar neighborhood \citep[e.g.][]{Wielen1977,Dehnen1998,Casagrande2011,Aumer2009,Bond2010,Smith2012,Binney2014b,Budenbender2015,Aumer2016}. As suggested by \citet{Budenbender2015}, these results might be affected by the tilt of the SVE, implying a coupling between the vertical and 
the radial velocity dispersion that is not taken into account in most  studies.

The aim of this paper is to revisit the SVE--Hubble type relation and try to 
reconcile the different results in the literature with the help of 
state-of-the-art simulations. We present, in \S\,\ref{sec2}, observational 
results for a sample of 55 galaxies gathered from published observational analyses and we describe the methods used to study their ellipsoids. 
Section~\ref{obsSVE} revises the SVE--Hubble type relation from the 
observational datasets presented here. 
In 
\S\,\ref{sec3} we introduce the set of numerical simulations used in this 
work. Section~\ref{sec5} makes use of the numerical 
work to elaborate on the sources of disc heating and 
their effect on the shape of the SVE. We sum up our conclusions in 
\S\,\ref{sec6}.

\section{Observational Data}
\label{sec2}

In this section, we describe the observed subsamples gathered from the 
literature. They are organized according to the methods used to calculate the 
shape of their SVE. Unfortunately, each subsample was analysed only with one method. 
So, we do not have the possibility of comparing the different techniques or verifying that they are consistent with each other. 
Nonetheless, we consider all the methods equally valid since they are all based on reasonable assumptions. 
We think that the general results of our study are robust to possible biases in individual methods.

\subsection{The solar neighborhood ellipsoid}
\label{MW}
Numerous works have been carried out for the SVE of the Milky Way disc in the solar 
neighborhood, with many of them leading to slightly different results depending on the specific sample. 
The study from \citet{Dehnen1998} involved stars in a wide range of populations, although representative only of the local solar neighborhood.
They selected a kinematically unbiased sample of 11\,865 stars from 
the \textit{Hipparcos} Catalogue ESA\,1997 \citep{Perryman1997}, which included also 
young stars. 
The \textit{Hipparcos} astrometry mission did not provide radial velocities for this big sample. 
They determined the kinematics from the 
absolute parallaxes and proper motions provided by this catalogue and using the 
deprojection technique described in \citet{Dehnen1998}. 
The velocity dispersion tensor was calculated in nine different bins in the $B-V$ color, all with equal number of stars. 
Their result was confirmed and updated by \citet{Aumer2009}, who used the new reduction of \textit{Hipparcos} data \citep{vanLeeuwen2007} and a 
27 per cent larger sample, with more blue stars. They re-calculated the velocity dispersion ratios as function of color and found a very good agreement 
also with the results from radial velocities from the Geneva-Copenhagen Survey \citep{Nordstrom2004}, in the range of red stars.

Some authors have recently extended the study of the stellar velocity ellipsoid of the Milky Way disc  to a larger volume of the solar neighborhood. 
\citet{Smith2012} made use of SDSS Stripe 82 proper motions \citep{Bramich2008} of 7280 dwarf stars, 
which they divided into three ranges in metallicity. They fitted the radial and vertical velocity distributions using maximum likelihood methods for 
different distances $|z|$ from the mid-plane (up to 2\,kpc). 
Other 16\,276 stars from SDSS, this time G-dwarfs from \textit{SEGUE} \citep{Yanny2009}, were used by \citet{Budenbender2015}. 
They constructed Jeans models to fit the orbital properties derived by \citet{Liu2012}. 
They studied the shape of the SVE in the meridional plane, between 0.5 and 3.0\,kpc away from the Galactic plane, for different metallicities.
\citet{Binney2014b} analysed the kinematics of $\sim$\,400\,000 stars from the RAVE survey \citep{Steinmetz2006}, in a region within $\sim$\,2\,kpc of 
the Sun.
 Probability density functions in distance modulus were determined by \citet{Binney2014a}. Proper motions were drawn from the UCAC4
 catalogue \citep{Zacharias2013}.
\citet{Binney2014b} mapped $\sigma_z/\sigma_R$ for different distances from the Galaxy plane.
 
In this context, light-weighted SVE measurements in external galaxies might be dominated by the young and more metal-rich thin disc. Nevertheless, 
the light is still integrated over the line of sight so that the thick disc has probably an important contribution 
(\citet{Yoachim2006} estimated the thick disc contribution to be between 10\% and 40\% of the total luminosity of the galaxy, depending on its mass).
Since we do not have any fair criterion to weight the different populations, we have decided to adopt, for the shape of the Milky Way disc ellipsoid, 
an average of the results of all the mentioned studies.
\looseness-2

\subsection{Velocity dispersions of edge-on galaxies: surface photometry}
\label{surfphot}
For edge-on galaxies, it is not possible to measure directly (with spectroscopy) 
the vertical component of the velocity dispersion. \citet{vanderKruit1999} 
proposed a method for its indirect estimate from surface photometry. They showed 
how to calculate the radial and vertical stellar velocity dispersions, with 
simple equations, knowing the radial scale length and the scale height of the disc. The 
following assumptions were necessary: exponential disc surface brightness radial 
profile, self-gravitating stellar disc (which can be approximated by an 
isothermal sheet), constant mass-to-light ratio ($M/L$) and flat rotation curves. 
They applied their model to 40 
edge-on late-type galaxies, 
most of them selected from the sample of 
\citet{deGrijs1998}, (e.g. inclination $i\ge 87^{\circ}$, angular blue diameters 
$R_{25}\ge 2.2$ arcmin, non-interacting, and discarding S0 and Sa because the 
assumptions needed in this method were not valid, in particular the self-gravitating disc).
The surface photometry and the structural 
parameters were already available, as well as HI observations which were used to estimate the gas-to-total disc mass.

They used the relations 
between the radial velocity dispersion, the radial scale length, the
vertical dispersion and the scale height as explained in 
\citet{vanderKruit1999}. 
The radial component of the velocity dispersion tensor ($\sigma_R$) was 
measured, at one photometric scale length in the $B-$band, using the empirical 
relations of \citet{Bottema1993}. The vertical component ($\sigma_z$), was computed (also at one scale length) assuming a Toomre parameter of 
$Q \approx$\,2 \citep{Toomre1964}. 
While they published the results by combining the 
individual measurements by morphological type, they kindly gave us access to 
values for 31 galaxies in the \citet{deGrijs1998} sample. 

\subsection{$\sigma_{LOS}$ decomposition from spectroscopic data}
\label{sigmaLOS}
After the Milky Way, the first object with a measurement of the 
vertical-to-radial velocity dispersion ratio was the Sb galaxy NGC\,488 
\citep{Gerssen1997}. They presented a new way of extracting the three components 
of the velocity ellipsoid, from long-slit spectroscopic observations of the 
line-of-sight velocity dispersion ($\sigma_{LOS}$) along the major and minor 
axes in intermediate-inclined discs. It used the epicycle approximation to 
relate $\sigma_{\phi}$ with $\sigma_R$ and the asymmetric drift equation 
to connect the circular velocity (described by a power-law) to the stellar 
rotation curve. Exponential radial profiles, with the same scale length, were 
assumed for both the radial and the vertical velocity dispersions. This implied 
that $\sigma_z/\sigma_r$ was assumed constant with radius in the disc-dominated region. 
The model had five free parameters which were fitted simultaneously 
to the major and the minor axes data. Since it was designed for the disc, it was 
applied only to the disc-dominated regions.\looseness-2 

Later, \citet{Gerssen2000} used the same procedure to study the SVE of the Sab 
NGC\,2985 (see also \citealt{Noordermeer2008}). The circular velocity, obtained from emission-line measurements in 
the gas disc, was included in the procedure. \citet{Shapiro2003} added four 
more galaxies and improved the previous results by \citet{Gerssen1997} and 
\citet{Gerssen2000} for NGC\,488 and NGC\,2985. Their study of the SVE along the 
Hubble sequence was completed with two later-type galaxies (NGC\,2280 and 
NGC\,3810) in \citet{Gerssen2012}. The same methodology was also applied by 
several authors to long-slit and integral-field observations \citep[see 
e.g.][]{Westfall2005,Westfall2011,Gentile2015}.

\subsection{Dynamical models: Schwarzschild, three integral and Jeans methods} 
\label{subsub:dyn}
Two-integral and three-integral models were used by \citet{Emsellem1999} to fit 
the photometry and kinematics of the S0 NGC\,3115.
Given that this galaxy has a double disc 
structure \citep[e.g.][]{Emsellem1999}, for the outer Freeman type II disc they 
constructed a three-integral model and recovered the $\sigma_z/\sigma_R$ ratio in the
meridional plane. In this paper, we use the mean value from $\sim$\,12\,kpc,  
where the outer disc starts to dominate the surface brightness.

\citet{Cappellari2006} built dynamical models, based on the 
\citet{Schwarzschild1979} numerical orbit-superposition method, for 24 
elliptical and lenticular galaxies part of the SAURON sample 
\citep{deZeeuw2002}. 
This subsample was selected because it had accurate distance determination, 
availability of \textit{HST} photometry, and did not present strong evidence of 
bars. The global anisotropy in the meridional plane was determined in 
\citet{Cappellari2007}, and parametrized with $\beta = 1 - \left( 
\sigma_z / \sigma_R \right)^2$. From the set of 24 galaxies, we include 8 lenticular galaxies in this work. 
The values of $\beta$ were calculated within one effective (half-light) radius, including 
not only the disc but also the central region of the galaxies. Following 
\citet{Gerssen2012}, we decided to include those results in our compilation, 
as the SAURON measurements should be biased towards the equatorial plane and 
also disc regions (given the extent of the aperture used).

\citet{Tempel2006} produced a model, based on Jeans equations, of the Sombrero galaxy (M\,104, NGC\,4594).
They used the averaged major-axis kinematics from previous literature works 
\citep{Kormendy1982,Carter1993,Hes1993,vanderMarel1994,Emsellem1996,Kormendy1996}
 and they determined the 
$\sigma_z$ and $\sigma_R$ radial profiles, as well as the orientation of the 
stellar velocity ellipsoid. In this paper, we adopt the mean value of 
$\sigma_z/\sigma_R$ in the disc-dominated range, between 2 and 10\,kpc approximately.
In a recent work, \citet{Kipper2016} studied the SVE of our neighbor galaxy M\,31. 
It is a well known galaxy and its major axis kinematics had been measured several times  
\citep{McElroy1983,Kormendy1988,vanderMarel1994,Kormendy1999}. The mass 
distribution was taken from \citet{Tamm2012}, who used a three-component model 
with a stellar bulge, a stellar disc and a dark matter halo. Jeans equations 
were used by \citet{Kipper2016} with an approach similar to \citet{Tempel2006}. 
The ellipsoid axis ratios were plotted in two different radial profiles for the bulge 
and the disc. We consider the mean value of $\sigma_z/\sigma_R$ in the disc for 
the purpose of this paper.

\section{The observed SVE - Hubble type relation}
\label{obsSVE}
As mentioned in \S\,\ref{sec1} and \S\,\ref{sec2}, several works in the literature 
\citep[e.g.][]{Dehnen1998,Merrifield2001,Gerssen2012,Aumer2009,Aumer2016} have suggested that the shape of the stellar 
velocity ellipsoid can be used to unveil different heating mechanisms. In 
addition, if the sources of disc heating were different for early and late-type 
galaxies, their SVE should have different shapes and show a trend along the 
Hubble sequence. We revisit these claims here by analysing all the observational 
results we could find in the literature on this issue. We present all these 
values of $\sigma_z/\sigma_R$ in Table~\ref{tab:obs}. 

\begin{figure}
\scalebox{0.9}{
\includegraphics[page={3},scale=0.28]{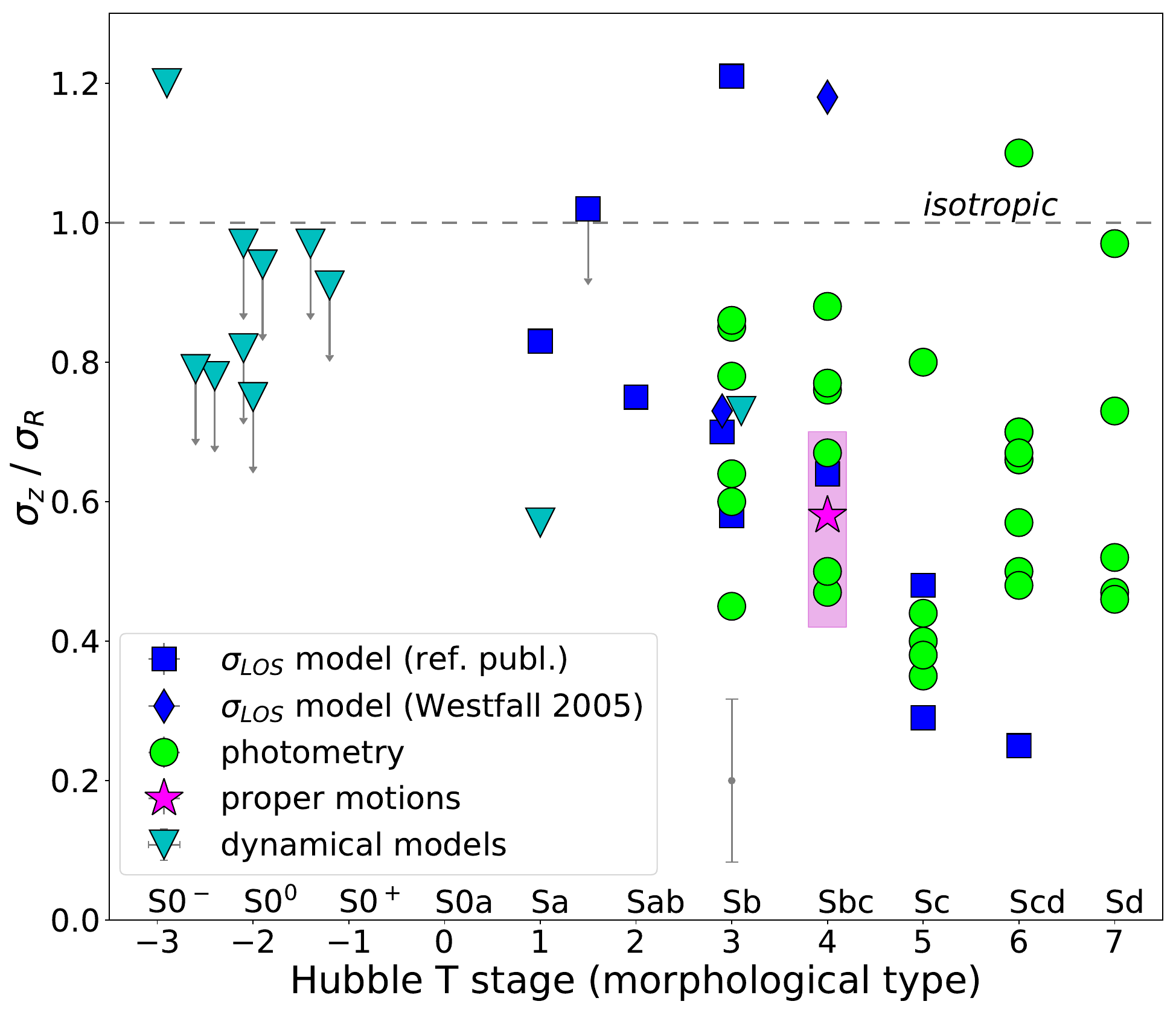}}
\caption{The flattening of the disc SVE, indicated by its axis ratio $\sigma_z/\sigma_R$, 
as function of Hubble types from observations of 55 galaxies. Different symbols 
and colors are used for different techniques and methods (see \S\,\ref{sec2} for 
details). \textit{Dark blue points} were computed from $\sigma_{LOS}$ models (squares for \citealt{Westfall2011,Gerssen2012,Gentile2015} 
and diamonds for \citealt{Westfall2005}, see \S\,\ref{sec2} for more details). 
In  \textit{green (circles)} results from surface 
photometry by \citet{vanderKruit1999}. 
In \textit{purple} the Milky Way in the 
solar neighborhood: the \textit{shade} shows the range where all the results mentioned in \S\,\ref{sec2} and \S\,\ref{obsSVE} lie, 
while the \textit{star} indicates their average value.
Finally, dynamical models (\textit{light 
blue triangles}) include Schwarzschild \citep{Cappellari2007}, three integral 
\citep{Emsellem1999} and Jeans models \citep{Tempel2006,Kipper2016}. 
Vertical arrows indicate upper limits. A mean error bar is indicated on the right of the legend. 
See Table~\ref{tab:obs} for more details.
 An horizontal line indicates the shape of the SVE when this is exactly isotropic ($\sigma_z = \sigma_R$).
} 
\label{fig:sra_H_obs}
\end{figure}

Figure~\ref{fig:sra_H_obs} shows all the literature $\sigma_z/\sigma_R$ values 
as a function of morphological type. Data points have been divided and 
represented with different colors and symbols according to the method used, 
as described in \S\,\ref{sec2}. The first thing to notice is that no clear correlation 
is found, opposed to the strong trend found by \citet{Gerssen2012}. 
A linear regression to all the points\footnote{Upper-limit values, as indicated in Table~\ref{tab:obs} and Figure~\ref{fig:sra_H_obs}, have not been taken into account.}
(weighting by their uncertainties) yields a slope
of $-0.03\pm0.02$, with a coefficient of determination ($R^2$, defined as the square of the correlation coefficient) of 0.15.
This indicates a very poor correlation between the axis ratio of the SVE and the 
morphological type. \citet{Gerssen2012} sample gives a slope of $-0.11\pm0.02$ with an $R^2$ of 0.86. 

The lack of a strong correlation is apparent in several individual datasets. The 
\citet{vanderKruit1999} sample is the largest and does not show any trend 
for their 31 galaxies, mostly very late types. One could argue that the (photometry based) technique 
employed is the cause for such discrepancy. 
Nonetheless, it uses approximations similar or common to other methods (e.g. exponential disc and isothermal sheet). 
Furthermore, the kinematics of 15 galaxies (11 in common with \citealt{vanderKruit1999}) was modeled, now using spectroscopy \citep{Kregel2004}, by \citet{Kregel2005a}. Their vertical-to-radial velocity dispersion ratios, plotted in Fig.\,6(a) of \citet{Kregel2005b}, 
range between $\sim$\,0.5 and $\sim$\,0.9 for Hubble types between Sb and Scd\footnote{They could not provide us the individual values of the plotted points nor the Hubble types they corresponded to. For this reason we have not been able to include them in Table~\ref{tab:obs} and Figure~\ref{fig:sra_H_obs}.}. 
Therefore, spectroscopic and photometric samples are compatible with each other and both point towards a picture where late types can have nearly isotropic SVE.

\citet{Westfall2005} and \citet{Gentile2015} data points were 
computed using the same methodology as \citet{Gerssen2012}, and yet they present 
very different values at a given morphological 
type. 
Two of these late-type galaxies display rather prolate ellipsoids (with 
$\sigma_{z}/\sigma_{r}>1$) suggesting that vertical agents must have been 
predominant. In their work, \citet{Gentile2015} excluded important recent
mergers as responsible for this heating, due to the regular morphology 
and kinematics. While these prolate ellipsoids may seem odd, there are several 
examples in our compilation, all of them coming from different sources and 
methodologies (see Table~\ref{tab:obs}). 
They might be related to a real phenomenon that we cannot explain yet, or be unreal and due to a general failure, common to the different methods, 
in reproducing the disc physics.
Even \citet{Gerssen2012} found a slightly vertically elongated ellipsoid for NGC2775, but considered the value of $\sigma_z/\sigma_R$ an upper limit.
The two galaxies from \citet{Westfall2005} come from an unrefereed publication and hence they are indicated with different symbols (diamonds) 
in Fig.~\ref{fig:sra_H_obs}. We decided to include them because they were obtained using the same $\sigma_{LOS}$ method as others and they are in 
agreement with others. 

For the Milky Way we indicate in Table~\ref{tab:mw} the individual results for $\sigma_z/\sigma_R$, from the different studies mentioned in \S\,\ref{MW}. 
They have been averaged over the different regions and ranges indicated in the third and the last columns of the Table, weighting with the number of stars in each bin or group when this was available. Errors have been estimated as the (weighted) standard deviations of the samples. 
\citet{Dehnen1998} believed that their value for the local solar neighborhood was the result of the contributions of both spiral structure and scattering by molecular clouds to the disc heating. 
For reference, \citet{Aumer2009} found $\sigma_z/\sigma_R$ from $\sim$\,0.33 for the bluest to $\sim$\,0.6 for the reddest stars in the \textit{Hipparcos} catalogue.
For the extended solar neighborhood, an increase in the kinematic isotropy for lower metallicities and larger distances $z$ from the Galactic plane was generally found.
\citet{Binney2014b} gave 0.6 as approximation for $\sigma_z/\sigma_R$ of giant and cool dwarf stars in the sample. For hot dwarfs, we have adopted the value given for the plane ($\sim$\,0.48) since the velocity dispersions increased slowly with distance from the plane. 

The range of values in Table~\ref{tab:mw} shows the lack of universal agreement on the global value of $\sigma_z/\sigma_R$ in the solar neighborhood. 
Different SVE shapes can be found even in the same galaxy, depending on the analysed populations. Results are sensitive to differences in color, 
metallicity, distance from the mid-plane and type of stars. 
We averaged the different results in Table~\ref{tab:mw}, weighting with the number of stars (in the second column of Table~\ref{tab:mw}), 
to find the value indicated in Table~\ref{tab:obs} for the Milky Way. The uncertainty has been estimated as the weighted standard deviation of the 
ensemble of samples. This value is represented with a purple star in Fig.\,\ref{fig:sra_H_obs}, where the purple shade shows the range of 
$\sigma_z/\sigma_R$ where the results in Table~\ref{tab:mw} lie.

\citet{Kipper2016} plotted the radial and vertical profiles of $\sigma_R$, 
separately for the bulge and for the disc of M\,31, showing that the bulge has 
radial velocity dispersions far larger than the disc. They also determined 
the orientation and axial ratios of the velocity ellipsoids of the two 
structural components along a meridional plane of the galaxy. While the 
ellipsoid stays almost spherical throughout the bulge component, the disc 
ellipsoid is nearly isotropic only near the rotation axis, and flattens out 
towards the most external radii. \looseness-2

For completeness, we also show the SVE values for the early-type galaxies in the 
SAURON survey \citep{Cappellari2007}. They have nearly isotropic ellipsoids, 
which is expected as they are the average within one effective radius ($R_e$), and thus 
include the bulge component. They should be considered as upper limits to the 
real $\sigma_z/\sigma_R$ of their discs. 

The main conclusion one can draw from Fig.\,\ref{fig:sra_H_obs} is that the 
relation between the shape of the SVE (using the $\sigma_z/\sigma_R$ ratio) and 
Hubble type is much weaker than previously reported. The earliest-type disc 
galaxies do not have necessarily isotropic ellipsoids and the latest types show 
a wide range of values. Therefore there must be different factors affecting the shape 
of the SVE of galaxies of the same morphological type. At the same time, 
different Hubble types can have the same ellipsoid shape. It is thus 
logical to wonder whether they are affected by the same heating mechanism or if 
different mechanisms can lead to the same $\sigma_z/\sigma_R$ ratios. However, since the region of late-type galaxies (in Fig.\,\ref{fig:sra_H_obs}) 
is completely dominated by the results from \citet{vanderKruit1999}, who used photometry, more results 
from spectroscopic observations would be necessary to confirm this picture.\looseness-2
\begin{figure}
\scalebox{0.9}{
\includegraphics[page={1},scale=0.28]{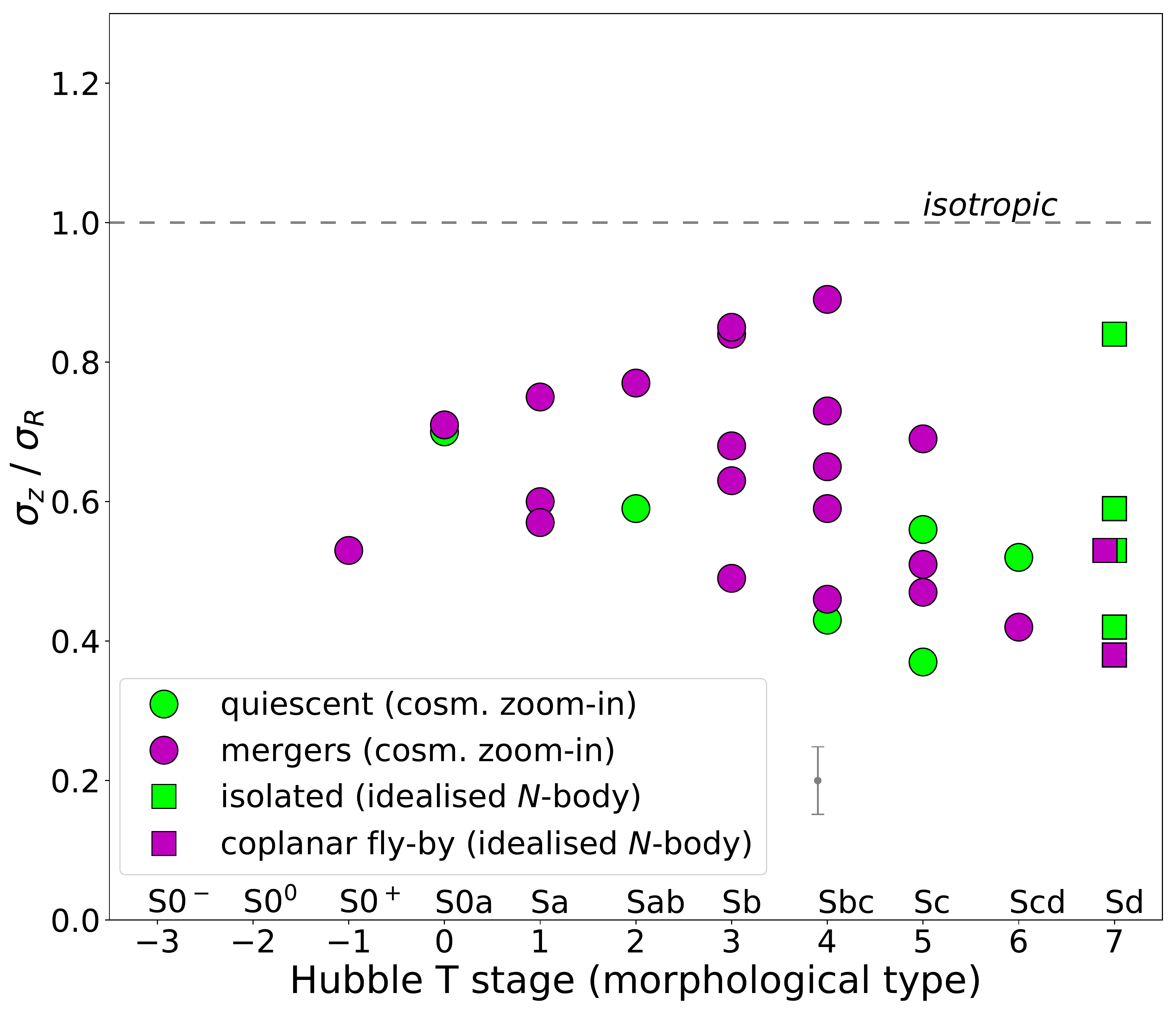}}
\caption{The flattening of the disc SVE, indicated by the ratio $\sigma_z/\sigma_R$, as 
function of Hubble types from simulations of 32 galaxies. The points have been 
divided with different colors and symbols. \textit{Light green circles} for quiescent (or merger with a mass ratio lower than 1:20)
and \textit{purple circles} for interacting galaxies in zoom-in 
cosmological simulations (up to 1:3 mass ratio mergers). 
This first classification was based on the galaxy evolution after its first 5\,Gyr of life. 
\textit{Purple squares} for idealised \textit{N}-body simulations 
with fly-by interactions (see \S\,\ref{sec3} for details). \textit{Green squares} come from idealised \textit{N}-body simulations that have evolved in 
complete isolation. 
The uppermost green square is a particular case in this sample (see \S\,\ref{sec3} and \ref{sec5} for details).
A mean error bar is indicated on the right of the legend. The individual errors were
computed as $\pm 1 \sigma$ of the disc radial distribution. See Tables~\ref{tab:zoom-in} and 
\ref{tab:idealised} for more details.
An horizontal line indicates the shape of the SVE when this is exactly isotropic ($\sigma_z = \sigma_R$).
} 
\label{fig:sra_H_sim}
\end{figure}
\section{Numerical simulations}
\label{sec3}
We describe here the two sets of numerical simulations used in this work. We use 
them to extract information of the SVE shape for comparison with the observational 
data and to explore the impact of different mechanisms.

\subsection{Disc galaxies from zoom-in cosmological simulations}
We determine the SVE information from the stellar kinematics of 26 simulated 
Milky-Way-mass spiral galaxies by \citet{Martig2012}, created with the 
purpose of studying galaxy evolution in a cosmological context. 
\citet{Martig2012} used a method that consisted in coupling cosmological
 with galactic scale simulations at much higher resolution. The 
technique is based on two steps: extracting the merger and gas accretion 
histories (and geometry) of dark matter halos in a large-scale $\Lambda$-CDM 
cosmological simulation, and then re-simulating these histories at very high 
resolution, replacing each halo by a realistic galaxy. 

The cosmological simulation was performed with the Adaptive Mesh Refinement code 
RAMSES \citep{Teyssier2002}. The simulation box had a comoving length of 20 
$h^{-1}$ Mpc and contained $512^3$ dark matter particles of $6.9\times 10^6$\,M$_{\sun}$ each. The cosmological constants in the $\Lambda$-CDM were 
set as 
follows: $\Omega_m$ = 0.3, $\Omega_{\Lambda}$ = 0.7, $H_0$ = 70 km s$^{-1}$ Mpc$^{-1}$ 
and $\sigma_8$ = 0.9. The merger and diffuse accretion histories of a halo were 
extracted by tracking halos and diffuse particles (not gravitationally bound 
with any halo). The position, velocity and spin of each incoming satellite were 
recorded as well as the date of the interaction. 

Each halo of the cosmological simulation (i.e. the main halo as well as all the 
interacting satellites) was then replaced with a realistic galaxy, made up of 
gas, stars and dark matter particles. The diffuse particles were replaced by a 
blob of gas and dark matter particles of lower mass and higher resolution. The 
total mass of the galaxy was divided in 20\% of baryons and 80\% of dark matter 
(the mass of dark matter being given by the cosmological simulation). The new 
simulation followed the evolution of the main galaxy from redshift $z$ = 5, included an area of 800\,kpc, to $z$ = 0. The re-simulation was 
done by means of the Particle-Mesh code described in \citet{Bournaud2002}, with 
a spatial resolution of 150 pc, a time step of 1.5 Myr and a particle mass of 
$1.5\times 10^4$\,M$_{\sun}$ for gas, $7.5\times 10^4$\,M$_{\sun}$ for stars and 
$3\times 10^5$\,M$_{\sun}$ for dark matter. A sticky-particle model was used for 
gas dynamics, and a Schmidt-Kennicutt law \citep{Kennicutt1998} for star 
formation, while supernovae feedback was introduced in the form of kinetic 
energy. In each galaxy substituting the dark matter halo from 
cosmological simulation, particles were already distributed in the morphological 
components. The disc, made of gas and star particles, was modelled with a Toomre 
profile \citep{Toomre1963}, while the bulge, made of stars, as a Plummer sphere \citep{Plummer1911}. A Burkert profile 
\citep{Burkert1995} was used for the dark matter halo.

Finally, the targeted galaxies were chosen to have a $z=0$ halo mass between $2.7\times 10^{11}$ and $2\times 10^{12}$\,M$_{\sun}$ and to be 
relatively isolated. They 
were distributed all across the cosmological simulation box, but avoiding the 
densest regions. They had different evolution histories, but most of them 
underwent mergers at some stage of their life-time. They have a wide range of 
bulge-to-total ($B/T$) ratios and morphological types.

\subsection{Idealised \textit{N}-body simulations}
In order to enrich our simulation sample with extremely late types and establish the influence of purely secular evolution processes in 
galaxies in total isolation, we also added to our study a small sample from 
idealised \textit{N}-body simulations by \citet{Martinez2006,Martinez2017}. They used the 
FTM-4.4 version of the \textit{N}-body code from \citet{Heller1994} and 
\citet{Heller1995}. The initial density distribution was derived from the 
disc-halo analytical model of \citet{Fall1980}. The disc was created 
exponential, with an initial scale length of 1.36\,kpc and a Toomre parameter 
fixed at 1.5. Regarding the velocity dispersion calculations, the initial 
$\sigma_R$ was set from the Toomre parameter and the surface brightness profile. 
The initial $\sigma_z/\sigma_R$ ratio was fixed to 0.6, a value commonly adopted in the literature
\citep[e.g.][]{vanderKruit1984, Bershady2011}, to 
define the initial $\sigma_z$.

We considered here a sample of six simulated galaxies. Four of them, identified by labels 
from I0 to I3, were allowed to evolve in isolation during the full simulation lifetime (14\,Gyr). All of them are bulgeless and 
were made with a disc and a dark matter halo, each one with $5\times 10^5$ 
particles at the beginning of the simulation. They differ from each other in the 
mass and dark matter fraction. The total mass decreases from I0 to I3,
in a range between $3.9\times10^{10}$\,M$_{\sun}$ (I3) and 
$2.0\times10^{11}$\,M$_{\sun}$ (I0). 
I0 contains 80\% of dark matter of the total mass budget within a 
sphere of radius 5\,kpc, while I1 contains 57\%. The I2 and I3 simulations have 
38\% and 7\% of dark matter fraction within 5\,kpc, respectively.

These simulations were created with the purpose of studying the mechanisms 
triggering the formation of bars, both in isolation and induced by interactions. 
For this reason, the two simulations I0 and I1 were redone but now with a galaxy fly-by and then 
left in isolation for $\sim$\,5\,Gyr. The interaction was modelled with the 
impulse approximation \citep[see][]{Martinez2017}. This means assuming that the 
energy exchanged during the interaction had been injected as kinetic energy to 
the host galaxy, whose reorganization within its potential happened after the 
fly-by had finished. The tidal approximation \citep{Binney1987} was also used. 
The perturber galaxy was assumed to have the same mass and size of the host and 
to orbit hyperbolically in the disc plane of the host. 

\subsection{Adequacy of numerical simulations}
The balance between the three-dimensional and the radial heating, the birth of new cooler young stars and how hot the older stars were born, determines the evolution of the stellar velocity ellipsoid shape. We have assessed these factors in our simulations, in order to assure ourselves that the results were adequately modelled.
While other authors opted for only GMCs as vertical heating source \citep[e.g.][]{Aumer2016}, in our zoom-in cosmological simulations we have instead a large variety of mergers, of different mass ratios and occurring in different planes and at different times.

As non-axisymmetric heating sources, bars in our idealised \textit{N}-body simulations match very well to real galaxies. 
Although not made for this purpose, \citet{Martinez2011,Martinez2013} and \citet{Gerhard2012} showed how one of these simulated galaxies (here called I1) 
reproduces the Milky Way bar and the corresponding boxy-bulge vertical metallicity gradient. 
These simulations have been also compared to nearby galaxies, by different authors. \citet{Seidel2015,Seidel2016} quantified the influence of bars on 
stellar kinematics and populations.  \citet{Molaeinezhad2016,Molaeinezhad2017} studied bulge properties and cylindrical rotation in barred galaxies.  
\citet{Font2017} analysed bar rotation and evolution. 
All of them found a good agreement between these simulations and observed galaxies.
\citet{Kraljic2012} studied bar formation and evolution in our zoom-in cosmological simulations, as well as the bar fraction in spiral galaxies, performing an analysis whose results were totally compatible with properties of real bars (e.g. length and strength).

A study on the effect of bars on the disc SVE shape was carried out by \citet{Martinez2006}, based on a disc simulation similar to our idealised 
\textit{N}-body simulations sample. In Fig.\,6 of their paper, the disc is affected very locally by the early evolution of the bar. 
A first bar growth just after its formation flattens the SVE by increasing $\sigma_R$. Then isotropy is rapidly reached thanks to the buckling 
(what they call "first buckling"). In our study, we take into account only the disc-dominated region, in order to perform an analysis as similar as 
possible to observational works. Therefore, these early effects on the inner disc do not leave any signature on our final SVE. 
Nevertheless, according to \citet{Martinez2006}, the later bar evolution does affect the kinematics of the outer disc. 
The final result comes from a trade-off between a potential secondary buckling, heating vertically, and the bar secular growth, heating radially. 
In our full simulations sample, no considerable effect of the buckling has been detected in the disc SVE, while an increase in $\sigma_R$ can be easily 
related to the secular growth of the bar.

Spiral arms, another non-axisymmetric feature affecting the SVE, have not been studied in our simulations yet. 
The measurement of their strength, or any other further analysis to quantify their impact on the disc heating, is beyond the scope of this paper.
However, similar dynamical processes dominate the formation of both bar and spiral arms. For this reason we expect spiral arms to be as realistic as bars in 
both zoom-in cosmological and idealised \textit{N}-body simulations.
Disc kinematics in our simulations are comparable to observations of real galaxies (as shown in 
Fig.\,\ref{fig:gal48} to \ref{fig:gal146} for zoom-in cosmological simulations). 
This points towards a good modelling of the disc heating. 
\citet{Martig2014b} compared the age-velocity relation (AVR) of some of their simulations with observations for the solar neighborhood, 
finding a very good match. 

The birth of new stars can counteract the heating by other sources, and affect $\sigma_z/\sigma_R$ if their ellipsoid shape is considerably different 
from the pre-existing populations. 
If the star-forming gas has cooled down and no new process has heated it, recently born stars are expected to be dynamically cooler than the ones born 
in the past. 
We have checked that new stars are born with lower velocity dispersions in the three directions, 
in our zoom-in cosmological simulations (see \S\,\ref{Time} for more details). 
In addition, \citet{leaman2017} have recently compared (in Fig.\,7 of their paper) the AVR in the Milky Way to one simulated galaxy from 
\citet{Martig2014b}. The latter not only matched the AVR in the solar neighborhood, but the velocity dispersion of its stars at birth, 
at different time steps, also matched 
the one of open clusters of different ages in the Milky Way disc.

However, very young stars in these simulations seem to be still too dynamically hot in their very first Myr of life, 
due mainly to the limited resolution and the gas density threshold for star formation, according to \citet{Martig2014b}.
Some tests confirmed that an increase in resolution or in the star formation threshold does not affect the shape of the AVR.
After the first 500 Myr, stars recover velocity dispersions similar to the Milky Way.
In addition, a different balance between the vertical and radial velocity dispersions, in recently born stars with respect to the rest of the galaxy, 
would be necessary to change the global $\sigma_z/\sigma_R$. For this reason, resolution and star formation recipe might not affect the shape of the SVE 
drastically.
Very young stars are expected to have almost circular orbits and fulfil the epicycle approximation, with a typical azimuthal-to-radial velocity dispersion 
ratio around  $0.7$ \citep{Binney1987}. This is true in these simulations and indicates that our modelling of the disc dynamics is reasonable.

In our idealised \textit{N}-body simulations, we do not have either mergers nor star formation. 
The purpose of including these simulations was to reproduce the behaviour of simple discs that are left to evolve by themselves. 
The only agents capable of modifying the shape of the SVE are the ones related to secular evolution, giving reasonable values of $\sigma_z/\sigma_R$ 
similar to zoom-in cosmological simulations and observations.

\section{Insights from numerical simulations}
\label{sec5}
We now turn our attention to the predictions from the numerical simulations
 to get a better insight into the mechanisms underlying the SVE of spiral galaxies.

\subsection{The simulated SVE - Hubble type relation}
We present in Fig.~\ref{fig:sra_H_sim} the 
vertical-to-radial velocity dispersion ratio as a function of the Hubble type 
for all the simulated galaxies. For zoom-in cosmological simulations, we 
assigned them a morphological type visually, by comparison with the observational sample classification. 
For all idealised 
\textit{N}-body simulations, given that they are bulgeless galaxies, we  
approximated their Hubble type to Sd, i.e. the latest type. 

The 
individual numerical values in 
Fig.\,\ref{fig:sra_H_sim} 
are shown in Tables~\ref{tab:zoom-in} and 
\ref{tab:idealised}. 
They are the medians computed within the disc-dominated region of the radial profiles, at the 
last time step of the simulations. 
The 
errors were computed as $\pm 1\sigma$ of the (radial) distribution. 
The disc radial range was defined as the one where the surface brightness profile could be fitted with an exponential. 
Thus, bulge and bar regions were excluded similarly to most observational results. 
The mean radial profiles for the individual 
dispersions were calculated by radially binning the star particles located at different azimuths. In this way, the effect of azimuthal variations 
was attenuated. However, we analysed these variations and verified that they were in general of the order of 5\,-\,10\%, although larger in the outer 
disc of some galaxies with pronounced spiral structure. 

Figure~\ref{fig:sra_H_sim} shows that no major trend is found with Hubble type. 
In these simulations, we have the possibility of looking for correlations between 
events or processes undergone by the galaxy and the shape of the SVE. For this reason, we have color-coded 
quiescent and non-quiescent galaxies, for both types of simulations. In zoom-in 
cosmological simulations none of the galaxies has evolved in total isolation. 
The galaxies plotted with light green circles are relatively quiescent, in the sense that 
they have not had 
mergers larger than 1:20 in mass of the satellite versus host galaxy, after the first 5\,Gyr of its lifetime (when we already had a stable disc).

No strong trend is seen globally for any given evolutionary path, although the light green circles suggest that quiescent galaxies SVE tend to be more 
isotropic for earlier types.
A linear regression to these six points gives a slope of $-0.04\pm0.03$ with an $R^2$ of 0.52, but more points would be needed to confirm this mild 
trend. Moreover, note that this relation is not observed in isolated galaxies in idealised \textit{N}-body simulations (green squares).
Like in observations, the lowest values of $\sigma_z/\sigma_R$ correspond to late-type spirals, which however show values in a wide axis ratio range. 
Quiescent and non-quiescent galaxies can have exactly the same ellipsoid shape.
This is in agreement with the study by \citet{Martig2012} on the same set of zoom-in cosmological simulations.
For disc-dominated galaxies, they found a correlation of the bulge-to-total light fraction not only with the merger history at $z<1$, but also with the 
gas accretion at $z>1$. 
More prominent bulges correspond to more active merger and/or gas accretion histories. 
Curiously, some of the bulges with the highest S\'ersic indexes were formed in absence of strong merger activity but with an early intense gas accretion 
together with an early bar formation or other disc instabilities. 
This suggests that there are different possible pathways to form a bulge, as 
\citet{Bell2017} confirmed observationally. In their sample, small pseudobulges were consistent with a quiet merger history and pointed to other formation 
mechanisms like gas accretion, disc instabilities or secular evolution. Some massive classical bulges were compatible with a merger origin, but others 
needed to invoke alternative mechanisms.
Therefore, no strong trend is expected between merger activity and morphological types, either.

While in observations, including early-type disc galaxies, 
the distribution of
$\sigma_z/\sigma_R$ has a slightly higher mean value of 0.7\,$\pm$\,0.2, 
these simulations suggest that the vast 
majority of spiral galaxies have a vertical-to-radial
axis ratio around 0.6\,$\pm$\,0.2. 
This would partly justify the 
common adoption of similar values in the literature 
\citep[e.g.][]{vanderKruit1984,Kregel2005b,Bershady2011,Martinsson2013}. It is important to 
remember that idealised \textit{N}-body simulations had this 0.6 as an initial 
condition, perhaps affecting somewhat the resulting ratios to be around this value, though with large scatter. 
On 
the contrary, no similar initial condition was set in zoom-in cosmological 
simulations. 
This simple result indicates that the vertical-to-radial 
axis ratio of the velocity ellipsoid appears to be a poor predictor of the 
dominant mechanism driving its shape. This finding is in agreement with 
the observational results presented in \S\,\ref{obsSVE}, and highlights the complexity of
interpreting the $\sigma_z/\sigma_R$ ratio.

Also in idealised \textit{N}-body simulations, discs evolved in total isolation have 
$\sigma_z/\sigma_R$ ratios extending in a wide range of values.
We have not found a direct correlation of the 
values with the dark matter fraction. A particular case is the isolated galaxy 
I0\_no\_bar (see Table~\ref{tab:idealised}) which, with an 80\% of dark matter in the central 5\,kpc, did not 
have the conditions to form any bar \citep{Martinez2017}. 
The cool disc of this galaxy, with low 
absolute values of the individual velocity dispersion components, has one 
of the most isotropic SVE in the full simulations sample.
Nonetheless, 
if this same disc undergoes a fly-by in the galaxy plane (with 1:1 mass ratio), 
this triggers the bar formation, causes a disturbance in the same plane which 
enhances $\sigma_R$, and flattens the ellipsoid (i.e. I0\_inter in 
Table~\ref{tab:idealised}). The same happens to the galaxy I1, where 
$\sigma_z/\sigma_R$ jumps down after the same interaction (I1\_inter). 

\subsection{Disentangling different processes from the SVE shape}

\begin{figure}
\scalebox{0.9}{	
\includegraphics[page={1},scale=0.9]{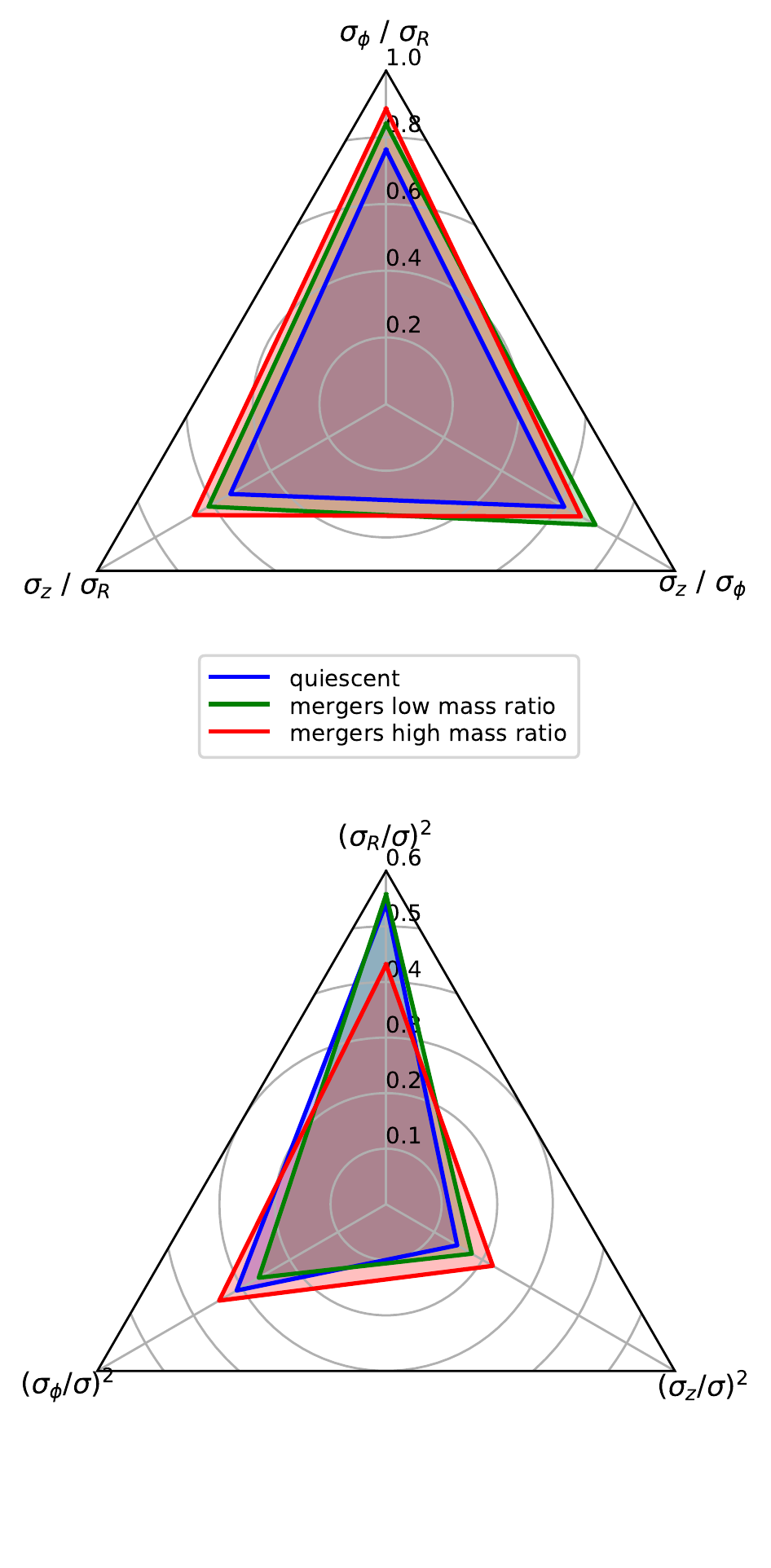}}
\caption{Radar plots of the average three axis ratios of the SVE (top) and 
normalized square velocity dispersions (bottom), for the zoom-in cosmological 
simulations sample, divided into three different groups according to their 
evolution after the first 5\,Gyr of life: 6 relatively quiescent galaxies, in \textit{blue} (no 
mergers with mass ratio higher than 1:20), 6 galaxies with ``low-mass'' mergers 
(i.e. 1:20 to 1:10) in \textit{green}, and 14 galaxies with ``high-mass'' mergers (i.e. 1:10 to 
1:3) in \textit{red}. $\sigma$ is the three-dimensional composition of the velocity 
dispersions defined as $\sigma^2$\,=\,$\sigma_{r}^2$\,+\,$\sigma_{\phi}^2$\,+\,$\sigma_{z}^2$.}
\label{fig:radar_merg}
\end{figure}

\begin{figure}
\scalebox{0.9}{	
\includegraphics[page={2},scale=0.9]{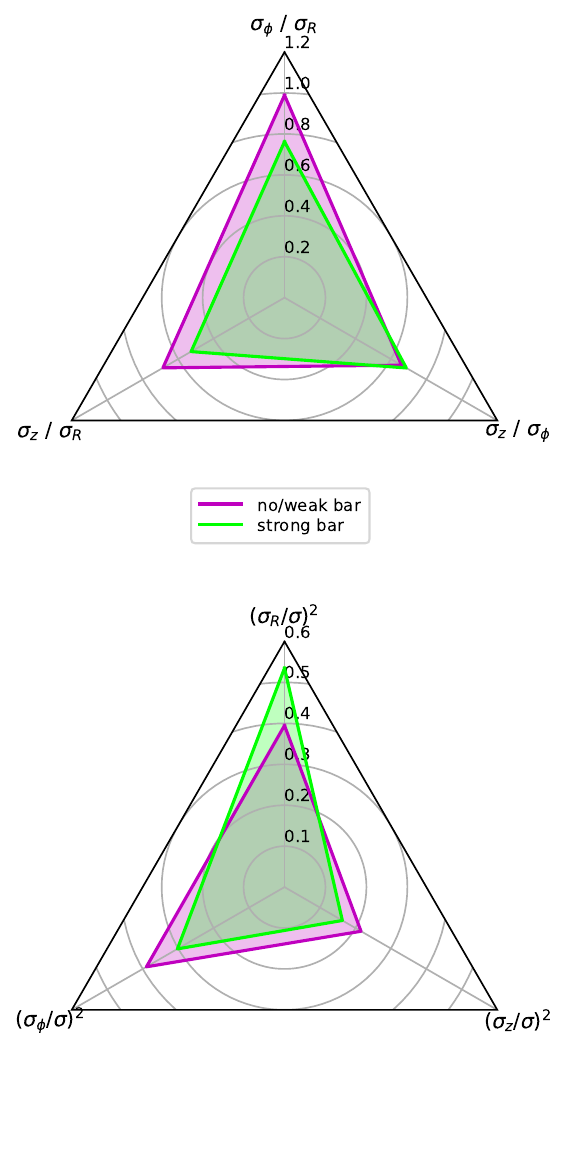}}
\caption{Radar plots of the average three axis ratios of the SVE (top) and 
normalized square velocity dispersions (bottom), for the zoom-in cosmological 
simulations sample, divided into barred and unbarred galaxies. 16 of them have a 
bar-to-total light ratio above 10\% (\textit{strong bars}) and 10 are below 6\%, 
having \textit{no or weak bars}. $\sigma$ is the three-dimensional composition of the 
velocity dispersions defined as $\sigma^2$\,=\,$\sigma_{r}^2$\,+\,$\sigma_{\phi}^2$\,+\,$\sigma_{z}^2$.}
\label{fig:radar_bars}
\end{figure}

Here we study in more detail the impact of some disc heating agents on the 
final shape of the SVE. This is shown in Fig.\,\ref{fig:radar_merg} and \ref{fig:radar_bars}, where the top 
panel shows the average axis ratios at the end stage of zoom-in cosmological simulations. 
The bottom panel presents the 
average of the square individual components, normalised by the square total dispersion defined 
as:  
$\sigma^2$\,=\,$\sigma_{r}^2$\,+\,$\sigma_{\phi}^2$\,+\,$\sigma_{z}^2$. In 
this way, it becomes clearer what is the most prominent axis of the SVE and helps us to
remove any potential dependence on the galaxy mass. Our zoom-in cosmological 
simulations allow us to specifically study the effect of mergers and also the 
action of bars. The effect of spiral arms is more complex to isolate, but will 
be discussed to some extent in the next section.

We focus first on the effect of mergers on the shape of the SVE (Fig.\,\ref{fig:radar_merg}). We divided the 
zoom-in cosmological simulations into three groups according to their evolution 
after the first 5\,Gyr of life, a time when in general the disc appearance starts to be similar to the final one. 
The first group (in blue) is made of relatively quiescent 
galaxies, with no mergers with mass ratios larger than 1:20 (i.e. light 
green points in Fig.\,\ref{fig:sra_H_sim}). The second group (in green) includes 
galaxies with ``low-mass'' mergers (i.e. 1:20 to 1:10 mass ratios). Finally, the third group 
is plotted in red and corresponds to the strongest mergers, with mass ratios 
larger than 1:10 (up to 1:3). 
The distributions of the second and third groups are not substantially different in Fig.\,\ref{fig:sra_H_sim}, 
hence we plot them together in purple - only their average values are somewhat different from one another, as shown in Fig.\,\ref{fig:radar_merg}.

We see in both panels that, on average, stellar velocity ellipsoids are more 
isotropic (i.e. axis ratios closer to 1 and individual components similar 
in amplitude to each other) when high-mass ratio mergers have taken place (red 
triangles). 
For $\sigma_z/\sigma_R$ in particular, high-mass mergers have a 
ratio $\sim$\,8\% larger than low-mass mergers. For 
the same axis ratio, a substantial difference is observed between galaxies with mergers and more quiescent galaxies. 
For the latter, $\sigma_z/\sigma_R$ is $\sim$\,14\% lower than for galaxies affected by 1:20-to-1:10-mass-ratio mergers.

The bottom panel shows that the (normalized) 
vertical component ($\sigma_z/\sigma$) is larger when high-mass-ratio encounters have 
occurred, and lower for more quiescent systems, where the radial component ($\sigma_R$) is more prominent 
(perhaps highlighting the prevalence of spiral arms or bars). 
Low-mass mergers can increase the contribution of $\sigma_z$ to the total dispersion, being an intermediate case
with respect to the high-mass mergers and quiescent galaxies. However, the contribution of $\sigma_R$ in these galaxies is similar to the quiescent group.
\looseness-2

To complete our analysis, we assess the influence of bars in 
Fig.\,\ref{fig:radar_bars}. Galaxies have been classified into two groups according to the 
bar-to-total light ratio, based on the photometric decomposition. 
Components with bar-to-total ratio above 10\%\ are considered strong 
bars and weak or no bars for values below 6\%. The figure suggests that bars (on 
average) are related to an enhancement of the radial component of the velocity 
dispersion. Consequently, galaxies with bars tend to have flatter ellipsoids 
(with $\sigma_z/\sigma_R$ and $\sigma_{\phi}/\sigma_R$ ratios smaller than 
in unbarred galaxies). The idealised 
\textit{N}-body simulations sample, with the only unbarred galaxy with a remarkably higher $\sigma_z/\sigma_R$, is also in agreement with this trend. 

While simulations allow us to see somewhat more clearly the effects of distinct 
mechanisms on the shape of the SVE, this is more difficult to capture in 
observations as differences are often subtle. In addition, different agents could have acted
simultaneously in real galaxies. The high scatter of the points even 
in groups with similar mergers history (presented in Fig.\,\ref{fig:sra_H_sim}) 
emphasizes this issue. Also, fly-by encounters can be important in enhancing one 
or another component of the velocity dispersion, as we have seen in idealised 
\textit{N}-body simulations. It is therefore likely that the particular 
dominance of one mechanism over others depends on the specific situations, which makes the shape of the SVE a not so obvious diagnostic to 
unveil the main mechanism influencing the dynamical heating of discs.
\begin{figure}
\flushright
\includegraphics[page={1},scale=0.201]{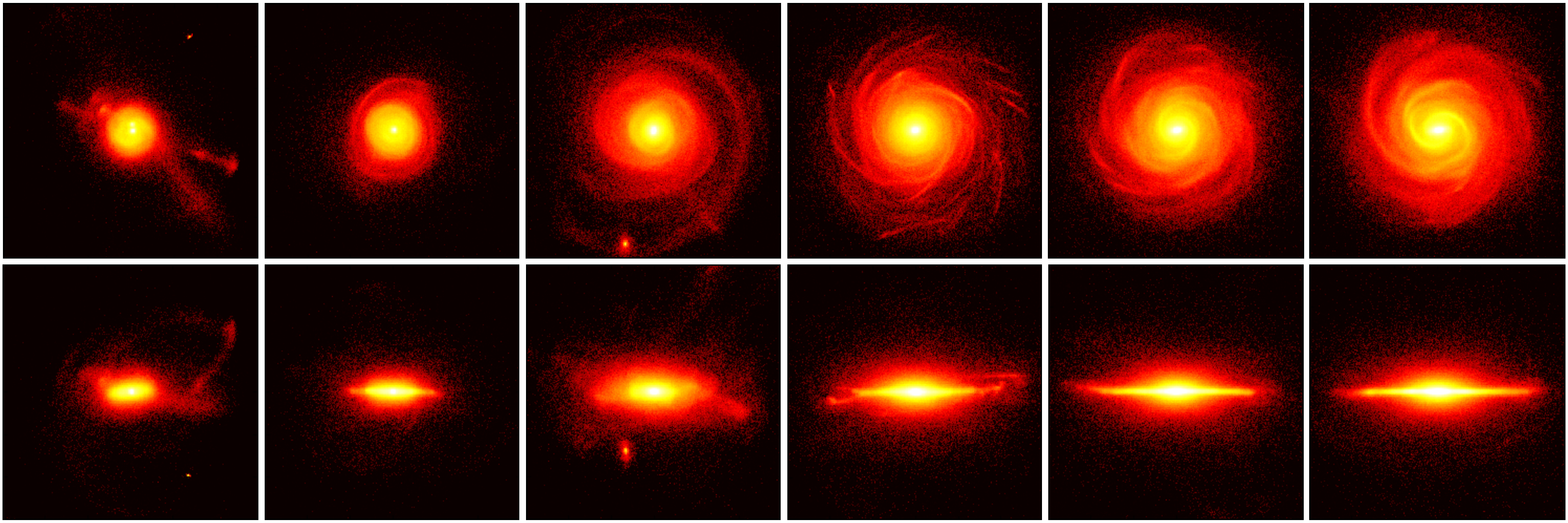}\\
\vspace{-5mm}
\flushleft
\includegraphics[page={1},scale=0.222]{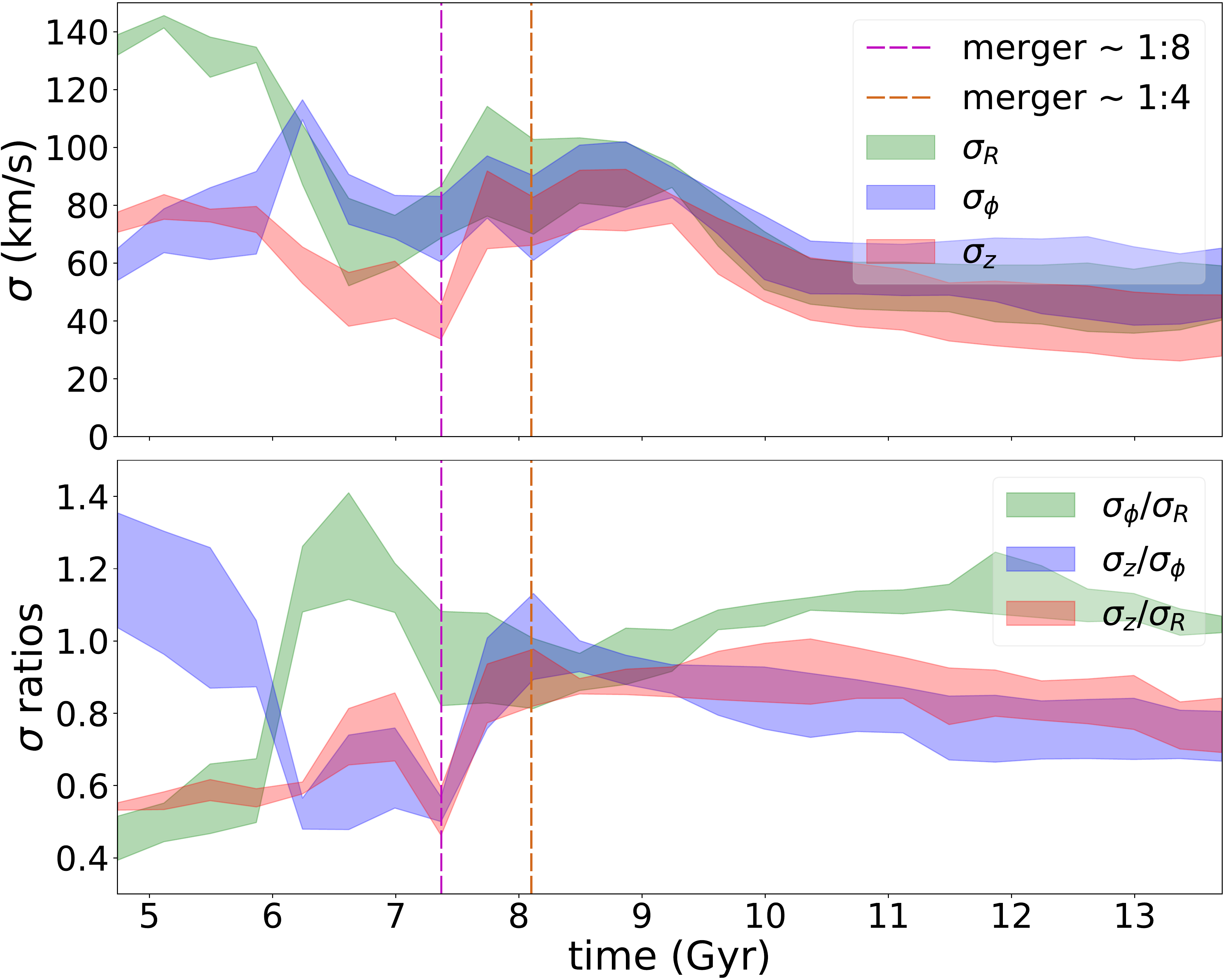}
\caption{Evolution in time of galaxy\_48: snapshots of stars (top panel, face-on 
in the first line, edge-on in the second line), stellar velocity dispersions (middle panel) 
and SVE axis ratios (bottom panel). 
The shaded areas mark the 16\%
and 84\% percentiles of the values calculated over the disc radial profiles (see \S\,\ref{Time} for details). 
Displayed on the horizontal axis is time since the Big Bang.
The galaxy had a 1:8 and a 1:4 merger around 
the ages of 7.3 and 8.1\,Gyr respectively, both indicated by vertical dashed 
lines.}
\label{fig:gal48}
\end{figure}
\begin{figure}
\flushright
\includegraphics[page={2},scale=0.201]{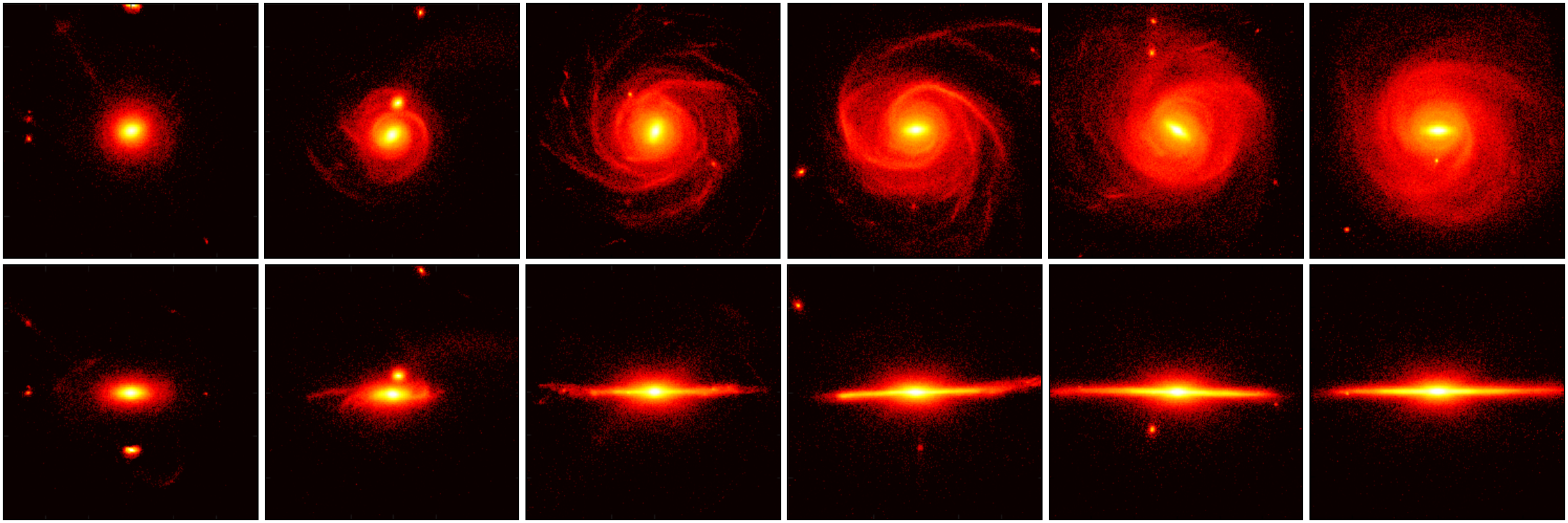}\\
\vspace{-5mm}
\flushleft
\includegraphics[page={2},scale=0.222]{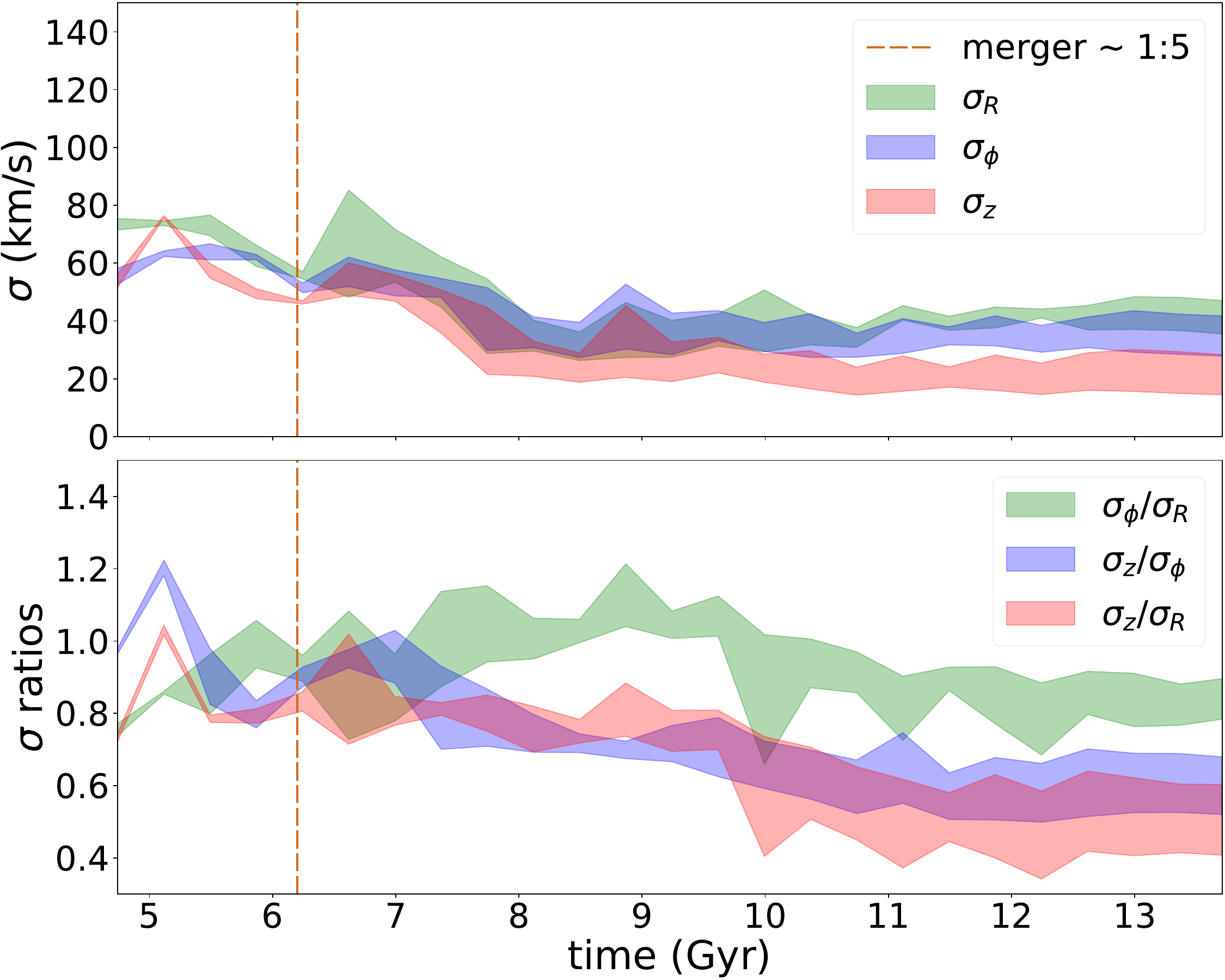}
\caption{Evolution in time of galaxy\_106: snapshots of stars (top panel, 
face-on in the first line, edge-on in the second line), stellar velocity dispersions 
(middle panel) and SVE axis ratios (bottom panel). 
The shaded areas mark the 16\%
and 84\% percentiles of the values calculated over the disc radial profiles (see \S\,\ref{Time} for details). 
Displayed on the horizontal axis is time since the Big Bang.
The galaxy had a long 1:5 merger from 
the age of 6 to 7\,Gyr, indicated by a vertical dashed line.}
\label{fig:gal106}
\end{figure}
\subsection{Time evolution of the SVE shape} \label{Time}

\begin{figure}
\flushright
\includegraphics[page={3},scale=0.201]{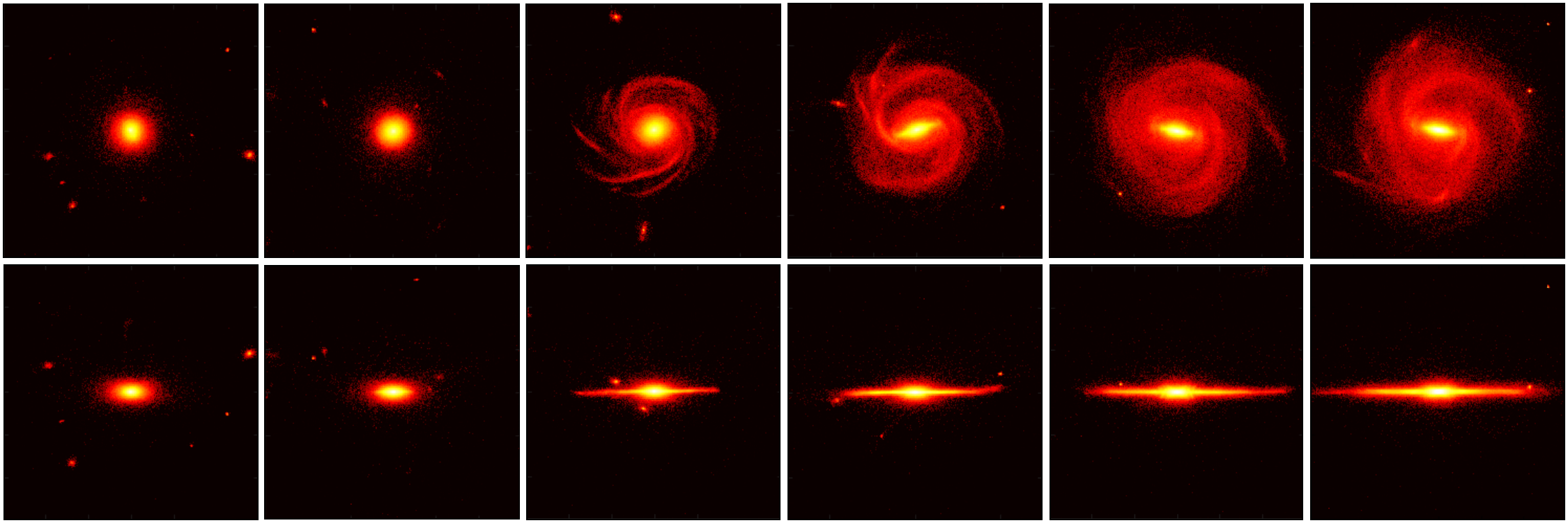}\\
\vspace{-5mm}
\flushleft
\includegraphics[page={3},scale=0.222]{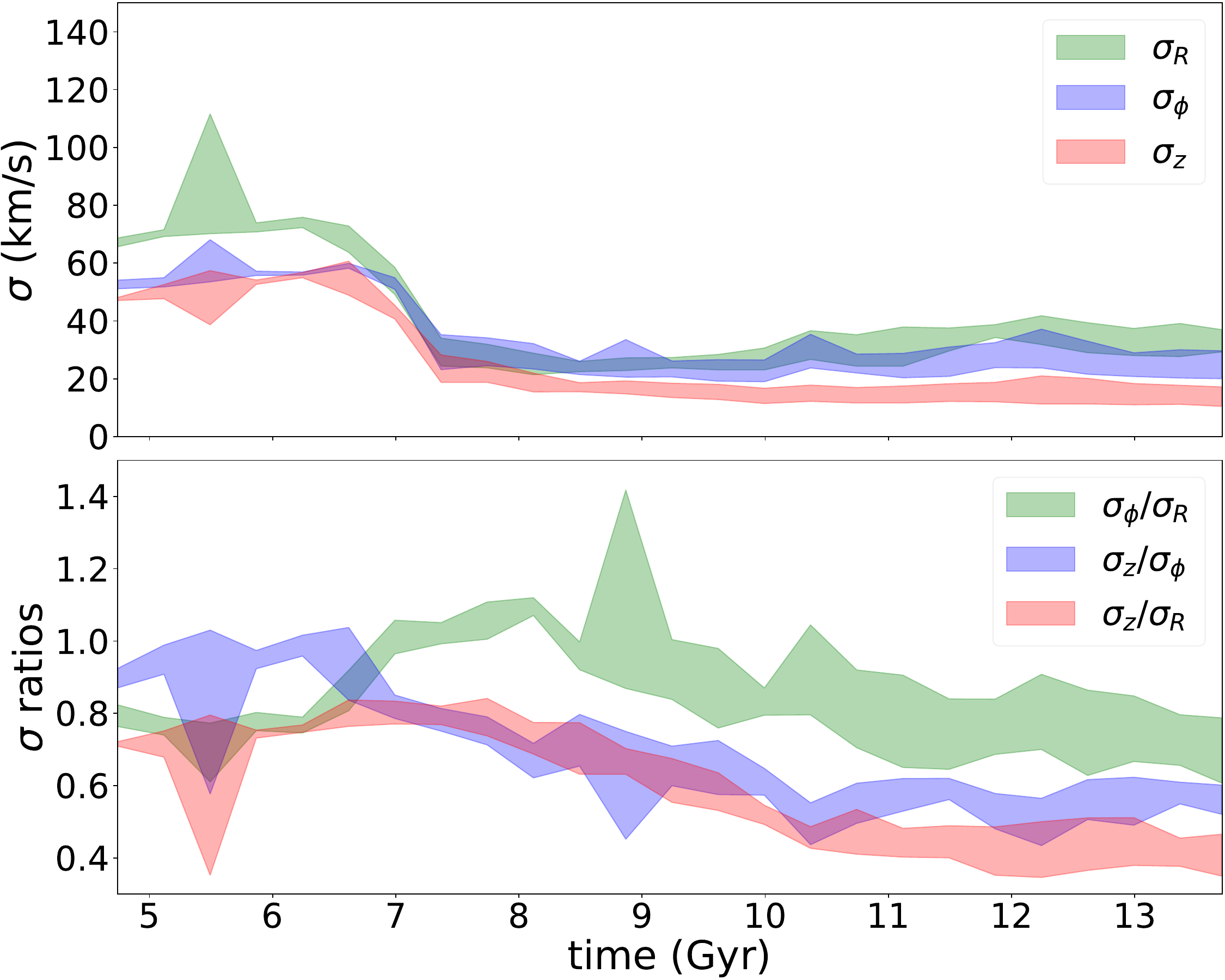}
\caption{Evolution in time of galaxy\_128: snapshots of stars (top panel, 
face-on in the first line, edge-on in the second line), stellar velocity dispersions 
(middle panel) and SVE axis ratios (bottom panel). 
The shaded areas mark the 16\%
and 84\% percentiles of the values calculated over the disc radial profiles (see \S\,\ref{Time} for details). 
Displayed on the horizontal axis is time since the Big Bang.
}
\label{fig:gal128}
\end{figure}

\begin{figure}
\flushright
\includegraphics[page={5},scale=0.201]{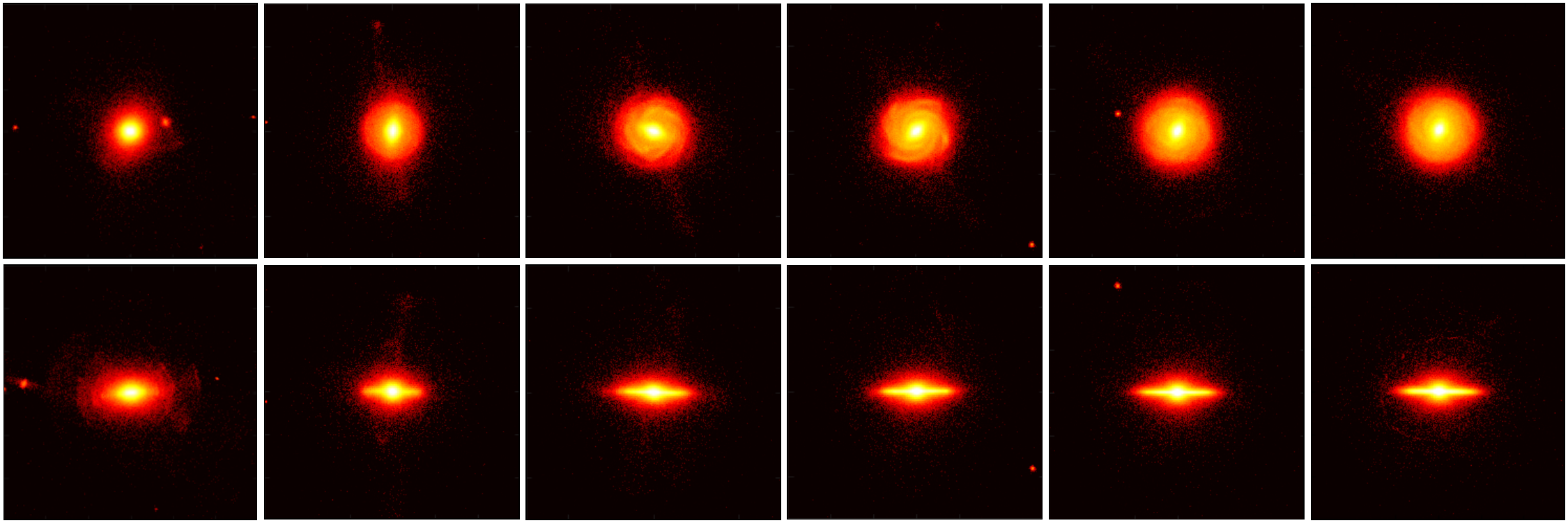}\\
\vspace{-5mm}
\flushleft
\includegraphics[page={5},scale=0.222]{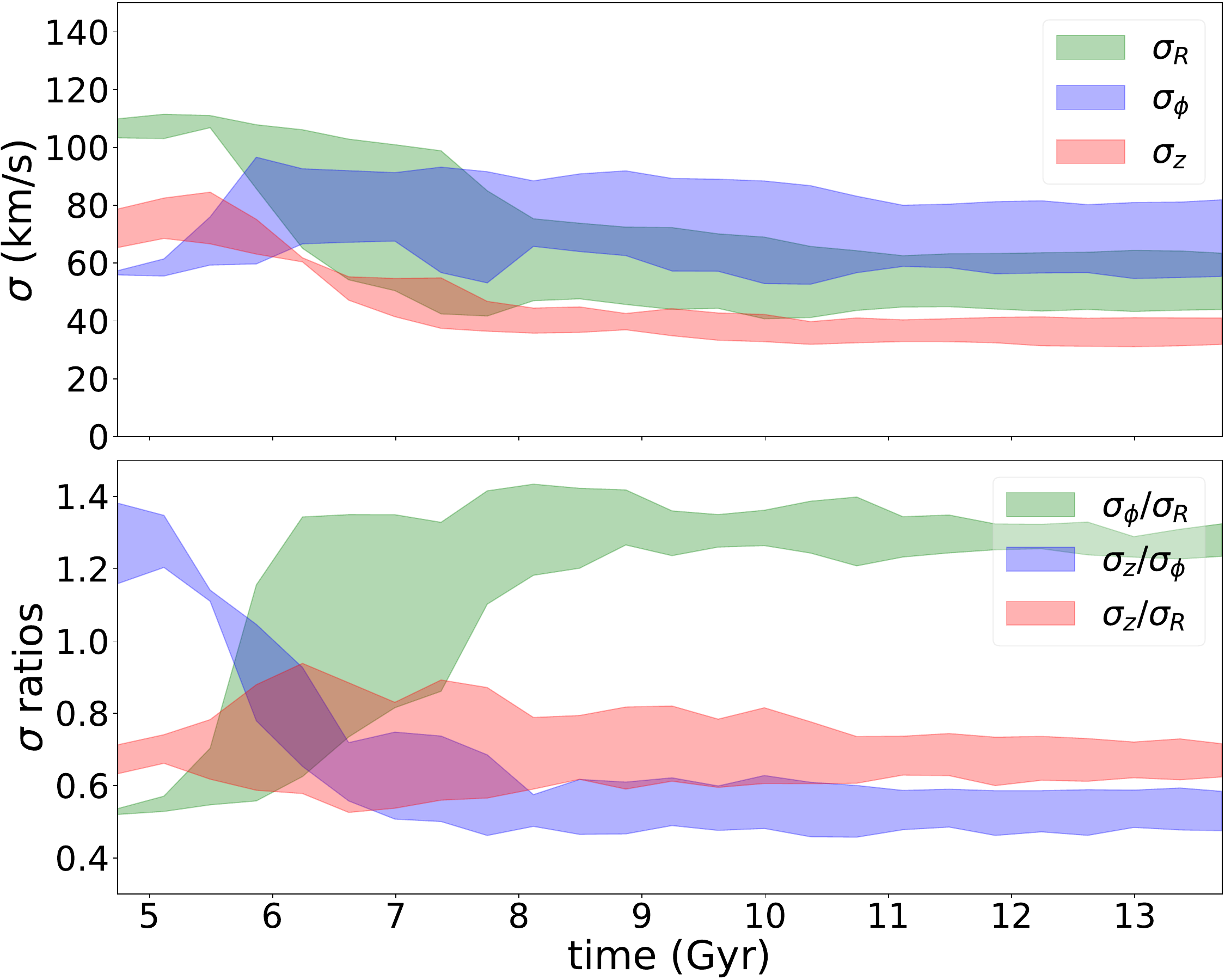}
\caption{Evolution in time of galaxy\_86: snapshots of stars (top panel, face-on 
in the first line, edge-on in the second line), stellar velocity dispersions (middle panel) 
and SVE axis ratios (bottom panel). 
The shaded areas mark the 16\%
and 84\% percentiles of the values calculated over the disc radial profiles (see \S\,\ref{Time} for details). 
Displayed on the horizontal axis is time since the Big Bang.
}
\label{fig:gal86}
\end{figure}

\begin{figure}
\flushright
\includegraphics[page={4},scale=0.201]{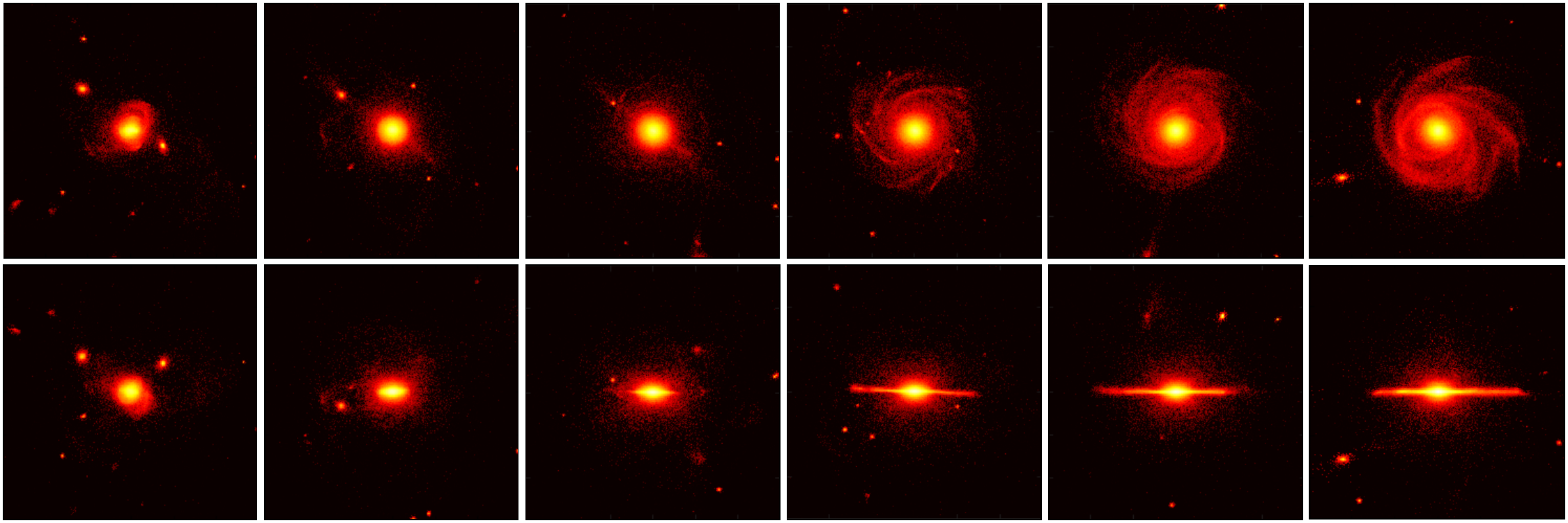}\\
\vspace{-5mm}
\flushleft
\includegraphics[page={4},scale=0.222]{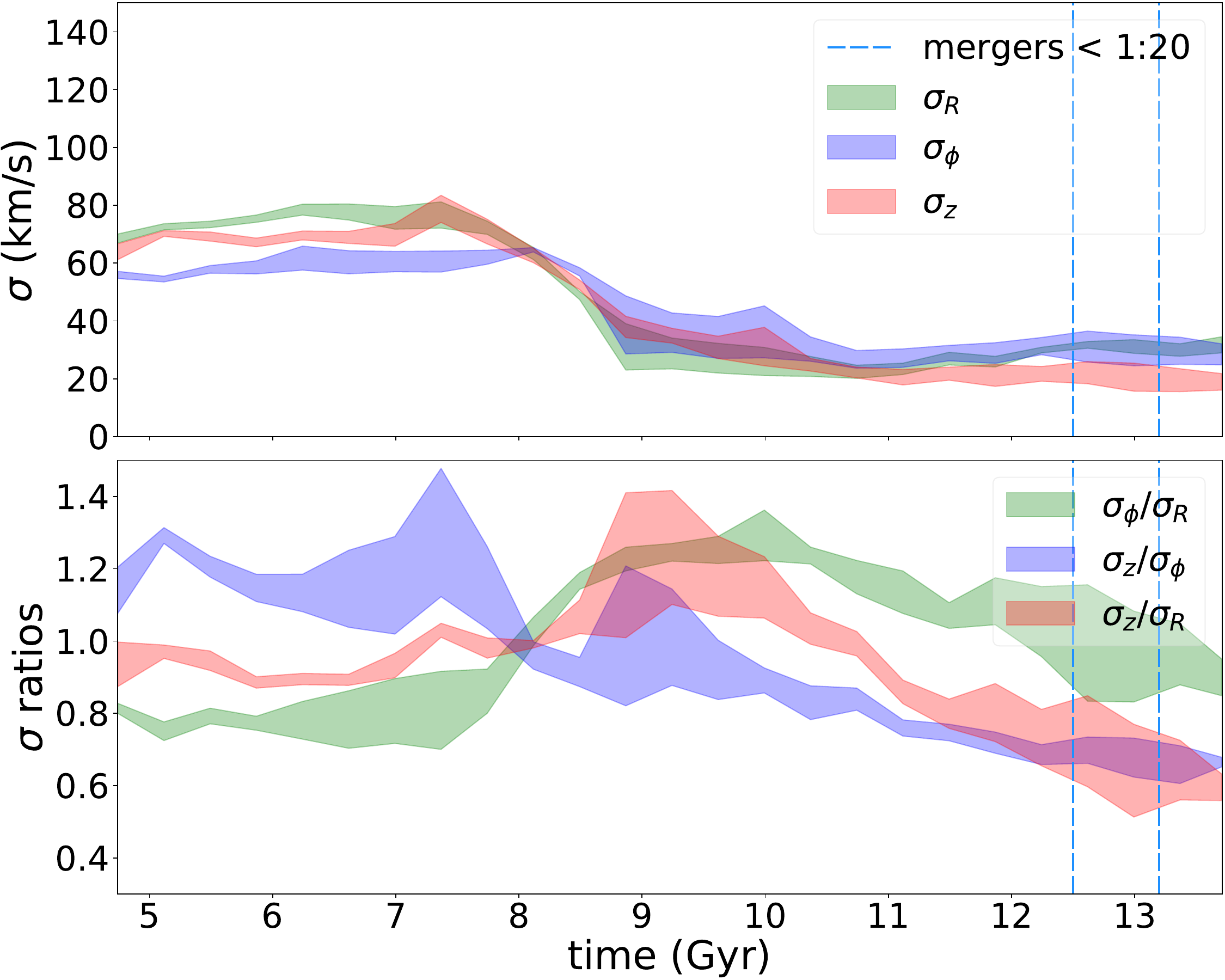}
\caption{Evolution in time of galaxy\_146: snapshots of stars (top panel, 
face-on in the first line, edge-on in the second line), stellar velocity dispersions 
(middle panel) and SVE axis ratios (bottom panel). Two mergers with very low mass 
ratios (<1:20) are indicated with dashed vertical lines. 
The shaded areas mark the 16\%
and 84\% percentiles of the values calculated over the disc radial profiles (see \S\,\ref{Time} for details). 
Displayed on the horizontal axis is time since the Big Bang.
}
\label{fig:gal146}
\end{figure}

Another interesting aspect to analyse in our simulations is the evolution of the 
shape of the SVE with time. This gives us information about how long a 
particular disc heating mechanism (e.g. mergers) influences the SVE shape. In 
Fig.\,\ref{fig:gal48} to \ref{fig:gal146} we plot the changes of the individual 
components and axis ratios of the SVE as a function of time, for five galaxies representative of the variety of cases. 
Of each group in Fig.\,\ref{fig:sra_H_sim}, we show one of the most and one of the least isotropic SVEs. Finally, 
we added one more quiescent galaxy as an example of continuously changing $\sigma_z/\sigma_R$. 
The shaded areas 
mark the 16\%\ and 84\%\ percentiles of the values calculated over the disc 
radial profiles. They are, at each specific time, the mean value in the radial profile and its 1\,$\sigma$ error bar. 
The radial profiles are the mean of all the star particles. 
Therefore these plots include, at a specific time, disc stars of all the previously born populations. In addition, we 
plot snapshots of the face-on and edge-on views of the simulated galaxies in 
order to determine the potential link between the shape of the SVE and galaxy 
morphology.

Mergers happen at different times in the life of different galaxies. In 
Fig.\,\ref{fig:gal48} we show the evolution of an Sbc galaxy (galaxy\_48). 
We see that the three dispersions, especially $\sigma_z$, after declining around the age of 6\,Gyr, 
increased steeply when the 
galaxy suffered from a 1:8 and a 1:4 mergers between the age of 7 and 8.5\,Gyr, indicated with 
vertical lines. Consequently, $\sigma_z/\sigma_R$ also increased to a value of 
$\sim$\,0.9. After the age of 9.2\,Gyr, the disc globally cooled down for little more than one Gyr and 
then the dispersions stayed approximately constant. In spite of this cooling, 
the vertical-to-radial ratio ended in $0.73_{-0.04}^{+0.12}$, much higher value than the initial ($\sim$\,0.55 at 4.7\,Gyr). 
This is not the case 
of galaxy\_106, a barred Sc type, with a final $\sigma_z/\sigma_R$ of 
$0.47^{+0.13}_{-0.06}$, much lower than the initial value ($\sim$\,0.8 at 4.7\,Gyr). 
Although this galaxy went through a long 1:5 merger from 6 to 7\,Gyr, 
it ended up with a quite flattened ellipsoid, as shown 
in Fig.\,\ref{fig:gal106}. The disc cooled down until 8\,Gyr, then dispersions 
stayed approximately constant.

\citet{Martig2014b} studied the effect of mergers on the age-velocity relation 
(AVR) of a sample including galaxy\_48 and galaxy\_106. In Fig.\,2 of their paper, both galaxies exhibit jumps in their AVR at the age 
corresponding to the end of the mergers. For galaxy\_48 the jump is more 
pronounced and the gas took much longer to cool down and give birth to cooler 
young populations. This is probably due to the fact that the mergers happened 
later, when the disc had already cooled down and new populations had a lower $\sigma_z$, being more sensitive. 
A small thin disc was also appearing, but during the merger epoch it became even thicker than before (see snapshots). 
In contrast, in galaxy\_106 the 
merger happened when the stellar disc was still thick and dynamically hot. In that time, internal as well as external perturbations 
had little effect so that 
the merger did not affect visibly the mean SVE shape. 
In addition, an almost imperceptible increase 
of the radial plus decrease of the vertical dispersions, was sufficient to make $\sigma_z/\sigma_R$ drop, between $\sim$\,9.5 and $\sim$\,11.5\,Gyr, 
from $\sim$\,0.8 to $\sim$\,0.37 (Fig.\,\ref{fig:gal106}). The origin of this gentle process could reside in the spiral density waves of this very 
late-type galaxy or the bar evolution, since it took place just after the spiral arms and the bar appeared (see snapshots in Fig.\,\ref{fig:gal106}). 
This picture provides a potential explanation for the differences in the 
SVE evolution of the two galaxies. 

We also show two examples of quiescence in Fig.\,\ref{fig:gal128} and 
\ref{fig:gal86}, galaxy\_128 being barred and galaxy\_86 unbarred. The first 
one, an Sc, exhibits the flattest ellipsoid in the sample (at the final step of the simulation). 
After a disc cooling already 
finished at $\sim$\,7.3\,Gyr, velocity dispersions remained approximately constant, if only 
with a slight increase of $\sigma_R$ over time. This was likely related to the development of the bar, the spiral structure 
and the thin disc (see snapshots in Fig.\,\ref{fig:gal128}). 
It was in fact enough to 
transform the ellipsoid from initially quite isotropic to much flatter, as seen 
in the bottom panel of the figure. The ellipsoid of galaxy\_86, one of the earliest types, 
on the contrary was quite isotropic from the start, 
and it barely changed during its quiescent path. 
Between its 6 and 8\,Gyr of life, the disc cooled down at the same rate in the vertical and the radial direction, but it heated remarkably in the 
azimuthal direction. This means that stars rotation became more disordered instead of ordered (as we have seen, discs typically cool down in the three 
directions meanwhile they acquire their final appearance). 
The peculiar properties of this disc are probably related to its morphology: it is photometrically smaller and thicker than galaxy\_128, 
with almost no spiral arms (see snapshots in Fig.\,\ref{fig:gal86}). 

Finally, in Fig.\,\ref{fig:gal146}, we show a relatively quiescent disc, 
born with an isotropic ellipsoid which became oblate and then flattened. During 
the disc cooling around 8\,Gyr, $\sigma_R$ dropped faster than $\sigma_z$, making 
the vertical-to-radial ratio larger than one. The snapshots indicate that the 
morphology of the galaxy drastically changed at that moment in time, from a small 
galaxy with no clear evidence of a disc to a fully grown spiral galaxy. From that 
time on, the $\sigma_z/\sigma_R$ ratio has decreased steadily to a value 
of $0.59^{+0.04}_{-0.03}$.

These examples lead us to some general deductions. 
First, all the discs seem 
to cool down before they acquire their final appearance (e.g. extended, fully 
formed and with spiral arms). 
The contribution of recently born stars can help us to understand the nature of this cooling. We have compared the kinematics of stars younger than 
300 Myr to the composition of all populations. 
The former have a degree of cooling that depends on the specific galaxy and time, if no mechanism heats dynamically the gas. 
During the mergers, they reach the velocity dispersions of the global galaxy, but afterwards they recover the previous values very soon.
Thus, a merger can delay the forthcoming cooling or make it more 
gentle, but cannot prevent it. 
Its signature can be erased by the appearance of new stars when enough time has passed.

The cooling observed in Fig.\,\ref{fig:gal48} to \ref{fig:gal146} was driven by the increasing contribution of new cooler stars. 
The latter were characterized by an earlier and faster drop (in around one Gyr) in the individual dispersions, during the formation of the thin disc. 
A delay was seen between the cooling of gas (new stars) and the cooling of the global galaxy stellar component.
An increase in the radial dispersion, driven by secular evolution, was also clearer in newly born stars 
compared to stars of all ages, where it was smoothed.
Despite this, the shape of the ellipsoid is not necessarily less isotropic for recently born populations. 
The shape comes from the balance between cooling or heating in the different directions, and therefore can be quite varied. 
For the two quiescent galaxies, galaxy\_128 and galaxy\_86, the same values of $\sigma_z/\sigma_R$ are measured for very young stars and for all 
populations. The remaining three galaxies analysed in this subsection have globally a more isotropic ellipsoid than recently born stars alone.

Secondly, 
the time in which we look at the 
ellipsoid is relevant. In some galaxies (e.g. galaxy\_48, galaxy\_106, 
galaxy\_128 and galaxy\_146) we would find an isotropic or anisotropic ellipsoid 
depending on when we look at it. 
Although at the final step of the simulations we do not have values of $\sigma_z/\sigma_R$ either larger than one 
or much smaller than 0.4, we do have them during the evolutionary path of some galaxies, 
in agreement with the observational results in those ranges.
Third, changes in the SVE shape are not 
necessarily related to disc heating, understood as an increase in velocity dispersions. These changes can also be caused by a decrease in radial 
or vertical dispersions, or an increase in both direction but in different 
amounts. 
Lastly, the vertical-to-radial ratio is more or less sensitive to 
small changes of $\sigma_R$ and $\sigma_z$ (few km s$^{-1}$), depending on their absolute values. 
Therefore in some very late types the slight increase in $\sigma_R$, contemporary to the appearance of the spiral structure, is sufficient to flatten 
the SVE but impossible to detect in observations and very sensitive to any other stronger process.
For these reasons $\sigma_z/\sigma_R$ might not be the best 
indicator of a disc heating or cooling, and even less helpful to distinguish between
individual heating agents.

\section{Conclusions}
\label{sec6}

We have gathered all the published studies analyzing the shape of the stellar velocity 
ellipsoid in galaxies. Aiming to extend the study done by \citet{Gerssen2012}, we have 
plotted the vertical-to-radial axis ratio as a function of Hubble type for 55 galaxies. No 
strong correlation has been found 
and 
the points are distributed with a large scatter around a mean value of $0.7\pm0.2$.

In general, the other works included in this paper are not in agreement 
with Gerssen and Shapiro's trend. 
This could be partially explained by the fact that their sample was carefully selected, as they explained \citep[e.g.][]{Shapiro2003}, 
for having "optically regular-looking morphologies" and the typical appearance of their specific Hubble type. Our sample includes 55 galaxies of all kinds and suggests a much more complicated picture.
Disc galaxies of early types can have more 
anisotropic ellipsoids than what their trend predicted, and late types can have 
ellipsoids covering all shapes. If it is true that different shapes correspond to 
different heating mechanisms, both early and late-type spirals can be affected 
by the same processes. At the same time, morphologically similar galaxies can be 
affected by different sources of disc heating. 

We have used two different kinds of simulations to understand better this 
observational scenario, and they confirmed this picture. The SVE's 
flattening does not show a clear trend along the Hubble type either in simulations. 
The average value is around 0.6 with a scatter 
of $\pm 0.2$. 
Idealised \textit{N}-body simulations confirm that late-type galaxies can have 
different kind of ellipsoids. Their shape does not show any direct relation with the 
distribution of the dark matter.
A relation between bars and the enhancement of the radial velocity dispersion is 
shown in zoom-in cosmological simulations.
According to idealised \textit{N}-body simulations, a fly-by in the disc plane also enhances $\sigma_r$ and flattens the SVE. Furthermore, 
1:10 to 1:3 
mergers in zoom-in cosmological simulations predict ellipsoids that are more 
isotropic on average, while quiescent mechanisms (perhaps bars and/or spiral 
arms) tend to enhance $\sigma_R$ with respect to the other components.
 
The scatter of $\sigma_z/\sigma_R$ is high also among galaxies with strong 
recent mergers and similar morphology. Evolution with time suggests that this 
axis ratio depends on the time when mergers happened and also on which 
snapshot of the galaxy life we are looking at. For instance, if just after a 
merger we see an isotropic SVE, its shape could change after some time.

In summary, the disc heating 
(cooling) scenario is complex and observations of the stellar velocity ellipsoid 
alone
are insufficient to unveil it, due to the following reasons:

\begin{itemize}
\item $\sigma_z/\sigma_R$ can be sensitive to small variations in one velocity 
dispersion component or two, perhaps not so relevant in terms of heating 
(cooling).
\item An isotropic ellipsoid can be related not only to an increase of the velocity dispersion in the vertical direction (heating), but 
also to a decrease in the radial direction (cooling), as well as to a static situation (if the SVE was born already isotropic). Likewise, anisotropy can 
be caused by a "radial heating" as well as a "vertical cooling".
\item When it comes to the different heating processes, the SVE results from the 
combination of them (e.g. mergers and spiral arms).
\item A given heating process can produce a variety of velocity ellipsoid shapes, depending on the time and the conditions in which it occurs.
\item Conversely, the same velocity ellipsoid can be achieved via a multitude of heating mechanisms.
\end{itemize}

The main outcome of this work is that additional information, beyond the stellar velocity ellipsoid, may be 
required to disentangle the heating agents in kinematic observations of galaxy 
discs. 
A potential way forward is to measure the velocity anisotropy separately
 for different stellar populations, to distinguish different mechanisms which happened at different moments along the life of a galaxy.

\section*{Acknowledgements}
We are grateful to Prof. Piet van der Kruit and Prof. Richard de 
Grijs for kindly providing the individual measurements presented in 
\citet{vanderKruit1999}. FP acknowledges Fundaci\'on La Caixa for the financial support received in the form of a Ph.D. contract. FP, JFB, GvdV and RL 
acknowledge support from grant 
AYA2016-77237-C3-1-P from the Spanish Ministry of Economy and Competitiveness 
(MINECO). JMA thanks support from MINECO through the grant AYA2013-43188-P. GvdV and RL were supported by Sonderforschungsbereich SFB 881
"The Milky Way System" (subproject A7 and A8) of
the Deutsche Forschungsgemeinschaft (DFG). GvdV acknowledges funding from the European Research Council (ERC) under the European Union's Horizon 2020 
research and innovation programme under grant agreement No 724857 (Consolidator Grant ArcheoDyn). RL acknowledges funding from the Natural Sciences and
Engineering Research Council of Canada PDF award.

\bibliographystyle{mnras}
\bibliography{biblio}

\begin{table*}
\centering
\caption{SVE shape for 55 galaxies calculated with different methods.}
\centering
\begin{tabular}{l|c|c|c|c|r} 
\hline\hline
\footnotesize
Galaxy name & Hubble type & $\sigma_z / \sigma_R$ $^{(1)}$  & References & Method & radial range\\
\hline\hline
Milky Way	 & Sbc $^{(2)}$ & $0.58 \pm 0.06$ & $^{(3)}$ & $^{(3)}$ & solar nbhd\\
\hline
ESO-LV 026-G06 & Sd & $0.52 \pm 0.08$ & \citet{vanderKruit1999} & photometry & at 1 scale length\\
ESO-LV 033-G22 & Sd & $0.47 \pm 0.07$ & \citet{vanderKruit1999} & photometry & at 1 scale length\\
ESO-LV 041-G09 & Scd & $0.70 \pm 0.10$ & \citet{vanderKruit1999}& photometry & at 1 scale length\\
ESO-LV 141-G27 & Scd & $0.66 \pm 0.10$ & \citet{vanderKruit1999}& photometry & at 1 scale length\\
ESO-LV 142-G24 & Sd & $0.73 \pm 0.11$ & \citet{vanderKruit1999}& photometry & at 1 scale length\\
ESO-LV 157-G18 & Scd & $0.66 \pm 0.10$ & \citet{vanderKruit1999} & photometry & at 1 scale length\\
ESO-LV 201-G22 & Sc  & $0.44 \pm 0.07$ & \citet{vanderKruit1999} & photometry & at 1 scale length\\
ESO-LV 202-G35 & Sb  & $0.60 \pm 0.09$ & \citet{vanderKruit1999}& photometry & at 1 scale length\\
ESO-LV 235-G53 & Sb  & $0.64 \pm 0.10$ & \citet{vanderKruit1999} & photometry & at 1 scale length\\
ESO-LV 240-G11 & Sc  & $0.40 \pm 0.06$ & \citet{vanderKruit1999}& photometry & at 1 scale length\\
ESO-LV 269-G15 & Scd & $0.70 \pm 0.10$ & \citet{vanderKruit1999} & photometry & at 1 scale length\\
ESO-LV 286-G18 & Sbc & $0.67 \pm 0.10$ & \citet{vanderKruit1999} & photometry & at 1 scale length\\
ESO-LV 288-G25 & Sbc & $0.88 \pm 0.13$ & \citet{vanderKruit1999} & photometry & at 1 scale length\\
ESO-LV 322-G87 & Sb  & $0.78 \pm 0.12$ & \citet{vanderKruit1999}& photometry & at 1 scale length\\
ESO-LV 340-G08 & Scd & $1.10 \pm 0.17$ & \citet{vanderKruit1999} & photometry & at 1 scale length\\
ESO-LV 340-G09 & Sd & $0.97 \pm 0.15$ & \citet{vanderKruit1999}& photometry & at 1 scale length\\
ESO-LV 383-G05 & Sbc & $0.76 \pm 0.11$ & \citet{vanderKruit1999} & photometry & at 1 scale length\\
ESO-LV 416-G25 & Sb  & $0.85 \pm 0.13$ & \citet{vanderKruit1999} & photometry & at 1 scale length\\
ESO-LV 435-G14 & Sc  & $0.80 \pm 0.12$ & \citet{vanderKruit1999} & photometry & at 1 scale length\\
ESO-LV 435-G25 & Sc  & $0.35 \pm 0.05$ & \citet{vanderKruit1999} & photometry & at 1 scale length\\
ESO-LV 435-G50 & Scd & $0.67 \pm 0.10$& \citet{vanderKruit1999} & photometry & at 1 scale length\\
ESO-LV 444-G21 & Scd & $0.57 \pm 0.09$ & \citet{vanderKruit1999} & photometry & at 1 scale length\\
ESO-LV 446-G18 & Sb  & $0.45 \pm 0.07$ & \citet{vanderKruit1999} & photometry & at 1 scale length\\
ESO-LV 446-G44 & Scd & $0.50 \pm 0.07$ & \citet{vanderKruit1999} & photometry & at 1 scale length\\
ESO-LV 460-G31 & Sc  & $0.38 \pm 0.06$ & \citet{vanderKruit1999} & photometry & at 1 scale length\\
ESO-LV 487-G02 & Sb  & $0.86 \pm 0.13$ & \citet{vanderKruit1999} & photometry & at 1 scale length\\
ESO-LV 505-G02 & Sd & $0.46 \pm 0.07$ & \citet{vanderKruit1999} & photometry & at 1 scale length\\
ESO-LV 506-G02 & Sbc & $0.47 \pm 0.07$ & \citet{vanderKruit1999}& photometry & at 1 scale length\\
ESO-LV 509-G19 & Sbc & $0.77 \pm 0.12$ & \citet{vanderKruit1999} & photometry & at 1 scale length\\
ESO-LV 531-G22 & Sbc & $0.50 \pm 0.07$ & \citet{vanderKruit1999} & photometry & at 1 scale length\\
ESO-LV 564-G27 & Scd & $0.48 \pm 0.07$ & \citet{vanderKruit1999} & photometry & at 1 scale length\\
\hline
NGC488  & Sb     & $0.70 \pm 0.16$ &\citet{Shapiro2003}&long-slit, $\sigma_{lOS}$ model&disc region\\
NGC1068 & Sb     & $0.58 \pm 0.07$ &\citet{Shapiro2003}&long-slit, $\sigma_{lOS}$ model&disc region\\
NGC2460 & Sa & $0.83 \pm 0.35$&\citet{Shapiro2003}&long-slit, $\sigma_{lOS}$ model&disc region\\
NGC2775 & Sa-Sab & $<1.02 \pm 0.11$&\citet{Shapiro2003}&long-slit, $\sigma_{lOS}$ model&disc region\\
NGC2985 & Sab    & $0.75 \pm 0.09$ &\citet{Shapiro2003}&long-slit, $\sigma_{lOS}$ model&disc region\\
NGC4030 & Sbc    & $0.64 \pm 0.28 $&\citet{Shapiro2003}&long-slit, $\sigma_{lOS}$ model&disc region\\
NGC2280 & Scd & $0.25 \pm 0.20$ &\citet{Gerssen2012}&long-slit, $\sigma_{lOS}$ model& disc region\\
NGC3810 & Sc  & $0.29 \pm 0.12 $& \citet{Gerssen2012}&long-slit, $\sigma_{lOS}$ model&disc region\\
NGC3223 & Sb   & $1.21 \pm 0.14$ & \citet{Gentile2015} &long-slit, $\sigma_{lOS}$ model&disc region\\
\hline
NGC3949 & Sbc  & $1.18 ^{+0.36}_{-0.28}$  & \citet{Westfall2005} & IFS, $\sigma_{lOS}$ model& disc region\\
NGC3982 & Sb   & $ 0.73 ^{+0.13}_{-0.11}$ & \citet{Westfall2005} & IFS, $\sigma_{lOS}$ model& disc region\\
NGC0234 & Sc   & $0.48 \pm 0.09$ & \citet{Westfall2011} & IFS, $\sigma_{lOS}$ model& disc region\\	
\hline
NGC524  & S0$^+$ & $<0.91 $ & \citet{Cappellari2007} & IFS, Schwarzschild models & within 1 $R_e^{(4)}$ \\
NGC3156 & S0     & $<0.78 $ & \citet{Cappellari2007} & IFS, Schwarzschild models & within 1 $R_e^{(4)}$ \\
NGC3414 & S0     & $<0.97 $& \citet{Cappellari2007} & IFS, Schwarzschild models & within 1 $R_e^{(4)}$\\
NGC4150 & S0$^0$ & $<0.82$& \citet{Cappellari2007} & IFS, Schwarzschild models & within 1 $R_e^{(4)}$ \\
NGC4459 & S0$^+$ & $<0.97$& \citet{Cappellari2007} & IFS, Schwarzschild models & within 1 $R_e^{(4)}$ \\
NGC4526 & S0$^0$ & $<0.94$& \citet{Cappellari2007} & IFS, Schwarzschild models & within 1 $R_e^{(4)}$\\
NGC4550 & S0$^0$ & $<0.75$& \citet{Cappellari2007} & IFS, Schwarzschild models & within 1 $R_e^{(4)}$\\
NGC7457 & S0$^-$ & $<0.79$& \citet{Cappellari2007} & IFS, Schwarzschild models & within 1 $R_e^{(4)}$\\
\hline
NGC3115 & S0$^-$ $^{(5)}$ & $1.2 \pm 0.1 $ & \citet{Emsellem1999} & Three-integral models & outer disc \\
\hline
M104    & Sa  & $0.57 \pm 0.12$ &\citet{Tempel2006}   & Jeans equations & disc \\
M31     & Sb $^{(5)}$ & $0.73 \pm 0.13$& \citet{Kipper2016}& Jeans equations & disc\\
\hline
\hline
\end{tabular}
\label{tab:obs}
\begin{tablenotes}
\item {\footnotesize Notes. (1) Uncertainties are indicated only when available.
(2) e.g. \texttt{http://messier.seds.org/more/mw\_type.html}. 
(3) See \S\,\ref{sec2} and \S\,\ref{obsSVE}. (4) Including bulge stars, these results are upper limits.
(5) \texttt{https://ned.ipac.caltech.edu}.\normalsize
}
\end{tablenotes}
\end{table*}
\begin{table*}
\centering
\caption{SVE shapes for the solar neighborhood.}
\centering
\begin{tabular}{l|c|c|c|c|r} 
\hline\hline
\footnotesize
 $\sigma_z / \sigma_R$ $^{(1)}$  & N$_{*}$ & Region & References & Comments\\
\hline\hline
$0.45\pm 0.06$ & 11865 & local solar nbhd & \citet{Dehnen1998}& $B-V$ from -0.238 to 0.719\\
\hline
{$0.42\pm 0.07$} & {15113} & local solar nbhd & \citet{Aumer2009}& $B-V$ from $\sim$\,0 to $\sim$\,0.8\\
\hline
$0.70\pm 0.07$ & 7280 & $|{z}|<2 $\,kpc & \citet{Smith2012}& dwarfs, [Fe/H]=[-1.5,0.5]\\
\hline
$0.58\pm0.04$ & 372\,768 & $|{z}|<2 $\,kpc & \citet{Binney2014b} & red-clump and non-clump giants, hot and cool dwarfs\\
\hline
$0.64\pm0.08$ & 16\,276 & $|{z}|<3 $\,kpc  & \citet{Budenbender2015}& G-dwarfs, [Fe/H]=[-0.89,-0.07]\\

\hline
\hline
\end{tabular}
\label{tab:mw}
\begin{tablenotes}
\item {\footnotesize Notes. (1) Averages have been computed weighting with the number of stars (N$_{*}$), in the second column. 
Errors have been estimated as the
weighted standard deviation of the different samples.\normalsize
}
\end{tablenotes}
\end{table*}
\centering
\begin{table*}
	\centering
	\caption{SVE shape for 26 galaxies from zoom-in cosmological simulations.}
		\centering
	\begin{tabular}{l|c|c|c|c|r} 
		\hline 		\hline
		\multicolumn{1}{c}{Galaxy} & Hubble  & Stellar mass           & Merger     & \multicolumn{1}{c}{\multirow{2}{*}{Bar$^{(2)}$}} & \multicolumn{1}{c}{\multirow{2}{*}{$\nicefrac{\sigma_z}{ \sigma_R}$}}\\
		\multicolumn{1}{c}{name}   & type    & ($10^{10}$ M$_{\sun}$) & mass ratio$^{(1)}$ & &\\
		\hline 		\hline
galaxy\_31  & Sbc & 1.2 & $1:10 - 1:20$ & strong & $0.46_{-0.00}^{+0.05}$\\
galaxy\_35  & S0$^{+}$ & 1.4 & $1:3 - 1:10$  & strong & $0.53_{-0.05}^{+0.17}$ \\
galaxy\_36  & Scd & 1.2 & $1:3 - 1:10$ & strong & $0.42_{-0.01}^{+0.05}$ \\
galaxy\_37  & Scd  & 1.2 & $<1:20$       & strong & $0.52_{-0.02}^{+0.03}$ \\
galaxy\_38  & Sc & 1.1 & $1:3 - 1:10$ & strong & $0.51_{-0.03}^{+0.06}$ \\
galaxy\_44  & Sbc & 1.2 & $1:3 - 1:10$  & strong & $0.59_{-0.04}^{+0.06}$ \\
galaxy\_47  & Sb  & 8.6 & $1:10 - 1:20$ & absent/weak & $0.68_{-0.02}^{+0.03}$ \\
galaxy\_48  & Sbc & 1.1 & $1:3 - 1:10$  & absent/weak & $0.73_{-0.04}^{+0.12}$ \\
galaxy\_53  & S0a & 9.1 & $1:3 - 1:10$       & strong & $0.71_{-0.04}^{+0.05}$ \\
galaxy\_56  & Sab & 1.1 & $1:3 - 1:10$       & absent/weak & $0.77_{-0.02}^{+0.09}$\\
galaxy\_57  & Sb & 1.2 & $1:10 - 1:20$        & strong & $0.49_{-0.07}^{+0.16}$ \\
galaxy\_59  & Sc & 7.2 & $1:3 - 1:10$  & strong & $0.69_{-0.04}^{+0.12}$ \\
galaxy\_60  & Sa & 4.5 & $1:3 - 1:10$  & absent/weak & $0.75_{-0.04}^{+0.02}$ \\
galaxy\_62  & Sb  & 6.7 & $1:10 - 1:20$ & absent/weak & $0.63\pm0.02$\\
galaxy\_72  & Sbc & 4.6 & $1:3 - 1:10$  & absent/weak & $0.89\pm0.02$       \\
galaxy\_82  & Sc & 3.8 & $<1:20$       & strong &$0.56_{-0.07}^{+0.04}$ \\
galaxy\_83  & Sb & 4.2 & $1:10 - 1:20$ & strong & $0.84_{-0.02}^{+0.05}$ \\
galaxy\_86  & S0a & 4.2 & $<1:20$       & absent/weak & $0.70_{-0.08}^{+0.03}$ \\
galaxy\_92  & Sbc  & 4.4 & $<1:20$       & strong & $0.43_{-0.06}^{+0.12}$ \\
galaxy\_102 & Sa & 3.3 & $1:10 - 1:20$ & absent/weak & $0.60_{-0.09}^{+0.07}$ \\
galaxy\_106 & Sc & 4.3 & $1:3 - 1:10$  & strong & $0.47_{-0.06}^{+0.13}$\\
galaxy\_125 & Sbc & 2.4 & $1:3 - 1:10$  & strong & $0.65_{-0.07}^{+0.03}$ \\
galaxy\_126 & Sa & 3.3 & $1:3 - 1:10$  & absent/weak & $0.57_{-0.04}^{+0.07}$ \\
galaxy\_128 & Sc & 2.7 & $<1:20$        & strong & $0.37_{-0.02}^{+0.09}$ \\
galaxy\_146 & Sab & 1.7 & $<1:20$        & absent/weak & $0.59_{-0.03}^{+0.04}$\\
galaxy\_147 & Sb & 2.6 & $1:3 - 1:10$  & strong & $0.85_{-0.01}^{+0.02}$  \\
	    \hline		\hline
			\end{tabular}	\label{tab:zoom-in}
		\begin{tablenotes}
				\item {Notes. (1) Estimated, for each galaxy, from its mass evolution plot from the age of 5\,Gyr on.
				(2) \textit{Strong} bars have a bar-to-total light ratio above 10\%. \textit{Absent/weak} bars are below 6\%.
				}
		\end{tablenotes}
\end{table*}

\begin{table*}
	\centering
	\caption{Axis ratio of the SVE for 6 galaxies from idealised \textit{N}-body simulations.}
		\centering
	\begin{tabular}{l|c|c|c|r} 
		\hline 		\hline
		Galaxy id & DM (\%)$^1$ & Bar & Interaction$^2$ & \multicolumn{1}{c}{$\sigma_z / \sigma_R$}\\
		\hline 		\hline
	    I0\_no\_bar & 80 & no & no  & \multicolumn{1}{c}{$0.84^{+0.16}_{-0.02}$} \\ 
	    I0\_inter  & 80 & yes & yes & \multicolumn{1}{c}{$0.53^{+0.26}_{-0.21}$} \\
	    I1        & 57 & yes & no  & \multicolumn{1}{c}{$0.59^{+0.08}_{-0.09}$} \\
	    I1\_inter  & 57 & yes & yes & \multicolumn{1}{c}{$0.38^{+0.16}_{-0.05}$} \\
	    I2        & 38 & yes & no  & \multicolumn{1}{c}{$0.42 \pm 0.09$} \\
	    I3        & 7 & yes & no  & \multicolumn{1}{c}{$0.53 \pm 0.08$} \\
	    \hline		\hline
			\end{tabular}	\label{tab:idealised}
		\begin{tablenotes}
		\item {Notes. (1) This is the dark matter fraction within a central sphere of 5\,kpc of radius.
		(2) The interaction consists in a fly-by, coplanar with the galaxy disc.
		}
		\end{tablenotes}
\end{table*}

\label{lastpage}
\end{document}